\documentclass[%
pre,preprint,
superscriptaddress,
tightenlines,
showpacs,showkeys,
twocolumn,
a4paper,12pt
]{revtex4}
\usepackage{dcolumn}
\usepackage{amssymb,amsmath,array}
\usepackage{graphicx}
\usepackage{subfigure}

\renewcommand{\thesubfigure}{(\alph{subfigure})}

\makeatletter
\renewcommand{\@thesubfigure}{\thesubfigure\space}
 
\newcommand{\bs}[1]{\boldsymbol{#1}}
\newcommand{\vc}[1]{\mathbf{#1}}

\newcommand{\dd}{\mathrm{d}}

\newcommand{\mrm}[1]{\mathrm{#1}}

\makeatother

\begin{document}
\DeclareGraphicsExtensions{.jpg,.pdf}
\title{Optical response of a nematic liquid crystal cell at splay--bend transition:
 model and dynamic simulation} 


\author{Peizhi Xu}
 \email[Email address: ]{pazixu@ust.hk}
 \affiliation{%
 Hong Kong University of Science and Technology,
 Clear Water Bay, Kowloon, Hong Kong
 }

 \author{Vladimir Chigrinov}
 \email[Email address: ]{eechigr@ust.hk}
 \affiliation{%
 Hong Kong University of Science and Technology,
 Clear Water Bay, Kowloon, Hong Kong
 }

\author{Alexei~D.~Kiselev}
\email[Email address: ]{kisel@mail.cn.ua}
\affiliation{%
 Chernigov State Technological University,
 Shevchenko Street 95,
 14027 Chernigov, Ukraine
} 

\date{\today}

\begin{abstract}
We study dynamical optical response of a nematic liquid crystal 
(NLC) cell that undergoes the splay-bend transition after applying the voltage
across the cell. We formulate a simplified model that takes into account both the
flexoelectric coupling and the surface rotational viscosity.
The dynamic equations of the model were solved numerically
to describe temporal evolution of the director profile and the
transmittance. We evaluate the response time as a function of
a number of parameters characterising
dielectric and elastic anisotropies,
asymmetry of the surface pretilt angles, anchoring energy,
surface rotational viscosity and flexoelectricity.

\end{abstract}

 
\maketitle

\section{Introduction}
\label{sec:intro}

As it was originally shown by Berreman and Heffner in 1981~\cite{Berrem:jap:1981},
a nematic liquid crystal (NLC) cell can be prepared to have two
metastable states that can be switched either way by applying
an electric field. This general idea underlies the mode of operation of
bistable liquid crystal devices that have been attracted considerable
attention over the past few decades.

The approach pioneered in~\cite{Berrem:jap:1981} is based on
using bistable twisted NLC cells that have two metastable twist states
produced as a result of a mismatch between the NLC pitch
and the twist imposed by the boundary conditions at 
the substrates. 
This approach has been extensively 
studied and is found to have difficulties caused by fast relaxation
of the metastable states to the intermediate stable 
configuration~\cite{Xie:jap:1998,Zhuang:apl:1999,Xie:jap2:2000,Xie:jap1:2000,Guo:apl:2000}.

An alternative approach is to use the so-called optically compensated
bend NLC cells also known as $\pi$
cells~\cite{Cheng:jap:1981,Acosta:lc:2000,Nakam:jjap:2000,Lee:jjap:2003,Inou:sid:2003}.
Boundary surfaces of such cells both favour a uniformly tilted
alignment and the pretilt angles at the substrates are equal in
magnitude but opposite in sign.
For sufficiently large surface pretilt angles, the equilibrium
orientational structures are
non-twisted~\cite{Port:jpp:1978,Cheng:jap:1981,Komit:pss:1986,Acosta:lc:2000}
and there are two director configurations that under certain
conditions are degenerate in energy:
the splay (horizontal) state and the bend (vertical) state.

By contrast to the bistable twisted cells, these bistable states
are topologically distinct and separated by an energy barrier.
So, the splay and bend states are both long-term stable.
Applying the voltage across the cell 
it can be switched from the splay state to the bend state.

This splay-bend transition will be of our primary concern.
We are aimed to study the dynamics of 
a NLC cell that undergoes the splay-bend transition
induced by an external electric field.

The dynamical theory of NLC systems~---~the so-called
nematohydrodymanics~---~is very complicated
and dynamical properties of bistable liquid crystal cells
have not received a fair amount of attention yet.
In recent theoretical studies
the dynamics of $\pi$ cells~\cite{Cheng:lc:2001}, 
zenithally bistable~\cite{Mottr:pre:2002} 
and super-twisted~\cite{Cheng:lc:2003} NLC devices
was investigated using different simplified models.

In this paper we concentrate on optical response 
of the NLC cell after switching on the voltage.
The corresponding response time will be studied depending
on a number of factors such as 
dielectric and elastic anisotropies,
asymmetry of the surface pretilt angles, anchoring strengths,
surface rotational viscosity and flexoelectricity.

The paper is organised as follows.

In Sec.~\ref{sec:model} we formulate our model
and derive a set of dynamic equations.
The numerical results are presented in
Sec.~\ref{sec:simulation-results}.
Concluding remarks are given in Sec.~\ref{sec:discussion}.

\section{Model}
\label{sec:model}

In this section we describe our model and
derive a set of dynamic equations. 
Subsequently, these equations will be
used to simulate the orientational dynamics of 
a NLC layer of the thickness $d$ that undergoes
the splay-bend transition under the action of an electric field.

\subsection{Free energy}
\label{subsec:free-energy}

The layer is sandwiched between two parallel plates, 
$z=0$ (lower substrate) and $z=d$ (upper substrate), and
we assume that both the electric field, $\vc{E}$, and 
the $z$--axis are normal to the plane of the substrates.
In addition, similar to~\cite{Cheng:lc:2001,Tsoy:2002,Mottr:pre:2002},
we shall restrict our consideration to the case in which
the splay-bend transition does not involve twisted 
states.

In this case, the NLC director field, $\vc{n}$,
is constrained to lie in the $x$-$z$ plane:
\begin{equation}
  \label{eq:director}
 \vc{n}=\cos\theta(z)\vc{e}_x + \sin\theta(z)\vc{e}_z, 
\end{equation}
where $\theta$ is the tilt angle
defined as the angle between the plane of the boundary surfaces and
the director; $\vc{e}_x$  and $\vc{e}_z$ are 
the unit vectors  parallel to the $x$--axis and the $z$--axis, respectively.

The vectors of easy orientation at the lower and the upper
substrates are similarly characterised by the tilt angles $\theta_L$ and
$-\theta_U$, respectively. So, the anchoring energy per unit area
taken in the Rapini-Papoular form~\cite{Rap:1969} is
\begin{equation}
  \label{eq:anch_energ}
  f_{\mrm{anch}}=\frac{W_{L}}{2}\,\sin^2(\theta_0-\theta_L)
+\frac{W_{U}}{2}\,\sin^2(\theta_1+\theta_U),
\end{equation}
where $\theta_{0,\,1}=\theta\bigr|_{z=0,\,d}$ and
$W_{L}$ ($W_U$) is the strength of anchoring at the lower (upper)
substrate.

We shall also need to write 
the bulk part of the free energy per unit area
\begin{equation}
  \label{eq:bulk_energ}
  F_b[\vc{n},\vc{E}]=
F_{el}[\vc{n}]+F_{E}[\vc{n},\vc{E}]
\end{equation}
which is a sum of the Frank elastic energy, $F_{el}[\vc{n}]$, 
and  the energy of interaction between
NLC molecules and the electric field, $F_{E}[\vc{n},\vc{E}]$.

For the director distribution~\eqref{eq:director},
using the standard expression for
the Frank elastic energy~\cite{Gennes:bk:1993} gives the following result:
\begin{equation}
  \label{eq:Frank_energ}
  F_{el}[\theta]=\frac{1}{2}\int_0^d
K_{el}(\theta)\,\dot{\theta}^2\dd z,
\end{equation}
where dot stands for the derivative with respect to $z$
and $K_{el}(\theta)=K_{11}\cos^2\theta+K_{33}\sin^2\theta$
is the effective angle-dependent elastic coefficient; 
$K_{11}$ and $K_{33}$ are the splay and the bend elastic constants.
Similarly, the director field~\eqref{eq:director} 
can be used to derive the expression for the electrostatic energy
$F_{E}[\vc{n},\vc{E}]$ that
depends on the electric field: $\vc{E}=E_z\vc{e}_z$.

Assuming that the NLC Debye screening length is larger than the layer
thickness, the NLC material can be regarded as an insulator.
So, we can neglect the effects caused by the presence of ionic charges.

But, the flexoelectric coupling between NLC
and the applied field cannot be generally disregarded. 
This coupling is known to be caused by 
splay and bend director distortions that give rise to 
an average flexoelectric polarisation
\begin{equation}
  \label{eq:flxpol_gen}
  \vc{P}_f=e_{11}\vc{n}(\bs{\nabla}\cdot\vc{n})+
e_{33}
(\vc{n}\cdot\bs{\nabla})\vc{n},
\end{equation}
characterised by the splay and bend flexoelectric coefficients,
$e_{11}$ and $e_{33}$.

This is the well-known flexoelectric effect, first described by Meyer
in 1969~\cite{Meyer:prl:1969}, which has been extensively studied over recent
years. Flexoelectricity appears to be a very important property of
NLCs which must be taken into account in all experiments that deal
with inhomogeneous director orientation. 

In our case, it is not difficult~\eqref{eq:Frank_energ} 
to obtain the $z$--component of
the flexoelectric polarisation~\eqref{eq:flxpol_gen}
in the following form:
\begin{equation}
\label{eq:flx_pol}
P_z = g(\theta)\,\dot{\theta},\quad
g(\theta)=e_f \sin\theta\cos\theta,
\end{equation}
where $e_f=e_{11}+e_{33}$ is the flexoelectric coefficient. 
So, the final result for the electrostatic energy is
\begin{align}
&
  \label{eq:E_energ}
  F_{E}[\theta,E_z]=
-\int_0^d\bigl[
\epsilon_{zz} E_z^2/2 + P_z E_z
\bigr]\dd z,
\\
&
\label{eq:eps_zz}
\epsilon_{zz}(\theta)=\epsilon_{\perp} (1+u\sin^2\theta),
\end{align}
where
$\epsilon_{ij}=\epsilon_{\perp}\delta_{ij}+(\epsilon_{\parallel}-\epsilon_{\perp})n_i
n_j$ is the dielectric tensor and 
$u = (\epsilon_{\parallel}-\epsilon_{\perp})/\epsilon_{\perp}$
is the dielectric anisotropy parameter.

The Maxwell equation $\bs{\nabla}\times\vc{E}=0$ implies that
the electric field $\vc{E}=E_z(z)\vc{e}_z$ 
can be expressed in terms of 
the scalar potential, $V$: $E_z=-\dot{V}$.
Variation of the electrostatic energy functional~\eqref{eq:E_energ} 
with respect to $V$ gives the well-known electrostatic 
constitutive relation
\begin{equation}
  \label{eq:constit_rel}
  -\frac{\delta F_{E}}{\delta E_z}=\epsilon_{zz} E_z + P_z = D_z,
\end{equation}
where $D_z$ is the $z$--component of the electric displacement field
that, in contrast to $E_z$, does not depend on $z$.

From the relation~\eqref{eq:constit_rel} 
the displacement $D_z$ can be expressed in terms of the
voltage $U=\int_0^d E_z\dd z = V(0)-V(d)$ as follows
\begin{align}
&
  \label{eq:D_z}
  D_z=\frac{U+\psi(\theta_1)-\psi(\theta_0)}{\int_0^d
  \epsilon_{zz}^{-1}(\theta)\dd z},
\\
\intertext{where}
&
\psi(\theta)=\int g(\theta)\,\epsilon_{zz}^{-1}(\theta)\,\dd\theta
\notag\\
&
=\frac{e_f \ln(1+u\sin^2\theta)}{2u\epsilon_{\perp}}.
 \label{eq:psi} 
\end{align}
The expression on the right hand side of Eq.~\eqref{eq:D_z}
clearly indicates the flexoelectricity-induced voltage shift.
The effects of this shift in optical response of hybrid aligned
liquid crystal cells were recently studied in~\cite{Kirk:lc:2003}.

Since the displacement $D_z$ does not vary across the layer,
it is convenient to have the displacement $D_z$ as an
independent field and use the free energy $G[\theta,D_z]$
which is related to the energy $F[\theta,E_z]$ via 
the Legendre transformation~\cite{Thur:jap:1981,Palier:prl:1986}
\begin{equation}
  \label{eq:Legendre}
  G[\theta,D_z]=F[\theta,E_z] + E_z D_z,
\end{equation}
where $E_z=(D_z-P_z)/\epsilon_{zz}$.
 
We can now combine Eqs.~\eqref{eq:anch_energ}--\eqref{eq:Frank_energ}
and Eq.~\eqref{eq:E_energ} to derive the free energy $G[\theta,D_z]$
in the following form:
\begin{align}
&
  \label{eq:F_Dz}
  G[\theta,D_z] = \int_0^d f_b\,\dd z + f_s,
\\
&
\label{eq:fb}
f_b = K(\theta)\dot{\theta}^2 + \frac{D_z^2}{\epsilon_{zz}(\theta)},
\\
&
\label{eq:fs}
f_s = f_{\mrm{anch}}+D_z (\psi(\theta_0)-\psi(\theta_1)),
\end{align}
where $K(\theta)=K_{el}(\theta)+g^2(\theta)/\epsilon_{zz}(\theta)$
is the effective elastic coefficient renormalised by the flexoelectricity.

As it can be seen from Eqs.~\eqref{eq:F_Dz}--\eqref{eq:fs},
the bulk elastic coefficient and the anchoring energy are both
renormalised by the flexoelectricity: $K_{el}\to K$ and 
$f_{\mrm{anch}}\to f_s$.
Static properties of NLC layers submitted to an electric field  are
known to be affected by this 
renormalisation~\cite{Dur:pra:1987,Dur:jap:1990,Zum:lc:1999,Mottr:pre:2003,Felc:lc:2003}.

\subsection{Dynamic equations}
\label{sec:dynamic-eqs}

Low-frequency dynamical properties of NLCs are generally characterised
by orientational relaxation as well as by shear and compressional
flow. A full set of dynamic equations governing 
nematohydrodynamics is known as the Ericksen-Leslie equations
and describes temporal evolution of the fluid velocity and 
 the director field.

When the characteristic time scale of the velocity field is much
shorter than the typical time of director reorientation,
the flow velocity can be adiabatically eliminated from the dynamics
of NLC. In this approximation, the orientational dynamics is
purely relaxational and can be formulated  as  a time-dependent 
Ginzburg-Landau model~\cite{Hohen:rmp:1977,Luben:bk:1995}.

We shall apply this model to obtain the dynamic equation governing 
the orientational relaxation of the tilt angle in the bulk.
Using the free energy~\eqref{eq:F_Dz} gives the following result
\begin{align}
&
  \label{eq:dyn_bulk}
  \gamma_b\frac{\partial\theta}{\partial t}=
-\frac{\delta G}{\delta \theta}= K (\theta)\ddot{\theta}+\frac{1}{2}\,\Bigl[K'(\theta)\,\dot{\theta}^2
\notag
\\
&
+[D_z/\epsilon_{zz}(\theta)]^2 \epsilon'_{zz}(\theta)\Bigr],
\end{align}
where $\gamma_b$ is the bulk rotational viscosity and 
prime stands for derivatives with respect to $\theta$.

It should be stressed that under certain circumstances
the backflow effect caused by the coupling between the fluid flow and 
the director may considerably affect dynamical characteristics of NLC 
cells. Specifically, the so-called ``optical bounce'' in twisted cells 
manifests itself as a dip in transmission of normally incident light 
after the electric field is 
turned off~\cite{Pier:jpp:1973,Berrem:jap:1975,Chen:pra:1991} .
But in cases where the twisted states are of minor importance
backflow is found to induce only quantitative changes in 
the dynamics ~\cite{Chen:pra:1991,Cheng:lc:2001} .

By analogy with Eq.~\eqref{eq:dyn_bulk} we can write the dynamic
equations for the tilt angles, $\theta_0$ and $\theta_1$,
at the lower and upper substrates
as follows~\cite{Mert:2000,Cop:pre:2001,Zih:pre:1996,Mottr:pre:2002}
\begin{equation}
  \label{eq:dyn_surf}
  \gamma_s \frac{\partial\theta_i}{\partial t}=
(-1)^i K(\theta_i)\dot{\theta}_i - \frac{\partial
  f_s}{\partial\theta_i},\: i=0,1,
\end{equation}
where $\gamma_s$ is the surface rotational viscosity, which is defined
as the ratio of the torque needed to change the director orientation
at the surface for a certain angle and the corresponding relaxation 
velocity~\cite{Dur:pre:2000,Dur:pre:1999}.

\begin{table}[htbp]
\begin{ruledtabular}
  \begin{tabular}{lc||lc}
$K_{11}$ (N) & 6.6$\times$10$^{-12}$ & 
$e_{11}$ (C/m) & -0.95$\times$10$^{-11}$
\\
$K_{33}/K_{11}$  & 3.0 & 
$e_{33}$ (C/m) &  -1.35$\times$10$^{-11}$
\\
$d$ ($\mu$m)  & 2.5 & 
$\epsilon_{\perp}$ & 6.3 
\\
$W$ (J/m$^2$)  &4.0$\times$10$^{-4}$  & 
$\epsilon_{\parallel}$  & 12.6 
\\
$\theta_L$  (deg) & 46.0 & 
$U$ (V) & 20.0 
\\
$\theta_U$  (deg) & 44.0 & 
$n_o$ & 1.5 
\\
$\gamma_b$ (N s/m$^2$) & 0.1 & 
$n_e$  & 1.6 
\\
$\gamma_s$  (N s/m) & 3.0$\times$10$^{-6}$ & 
$\lambda$ ($\mu$m) & 0.55 
\\
  \end{tabular}
\end{ruledtabular}
  \caption{Parameters of the model employed in the calculations.}
  \label{tab:params}
\end{table}

\begin{figure*}[!tbh]
\centering
\resizebox{150mm}{!}{\includegraphics*{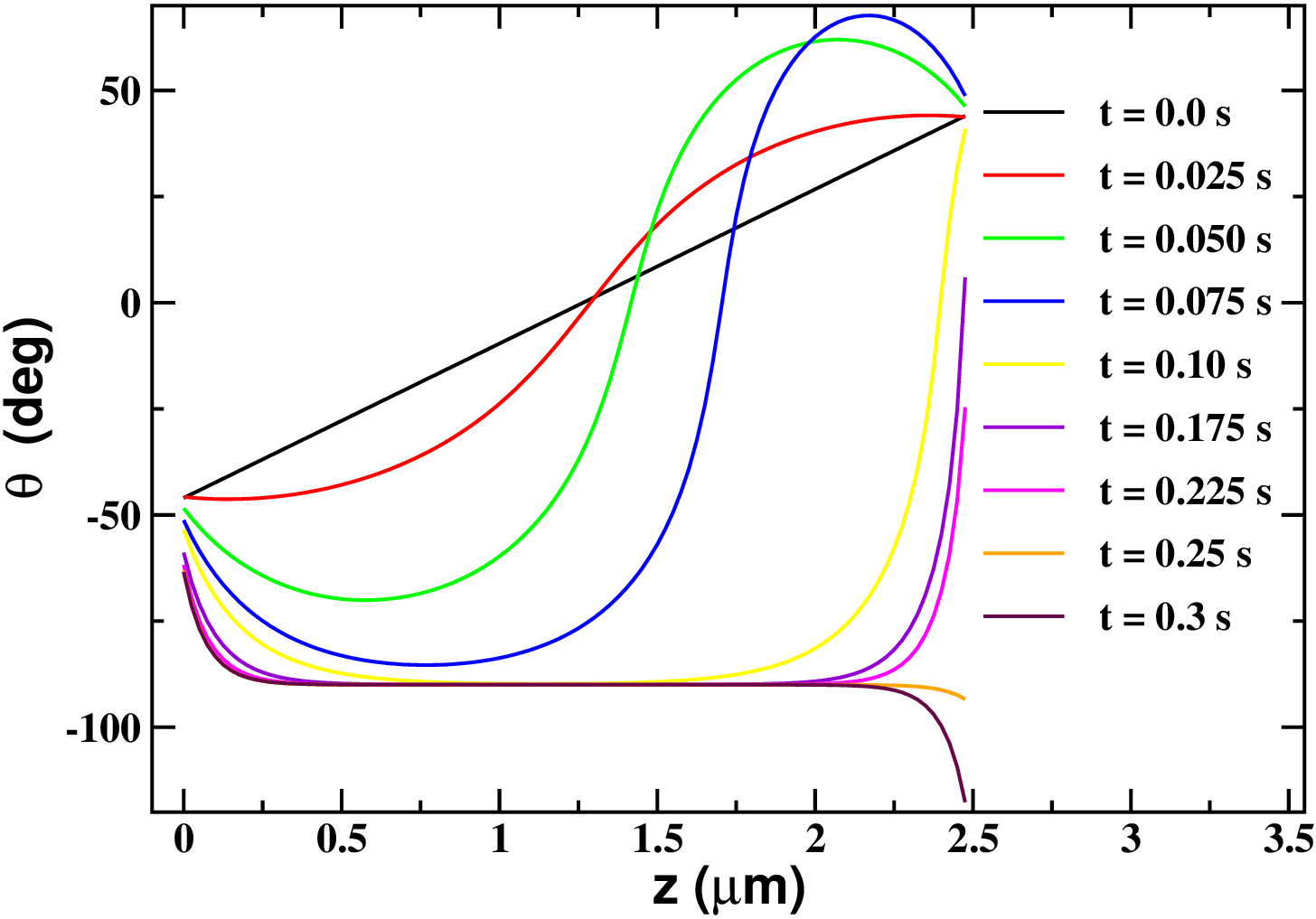}}
\caption{%
The director configuration through the cell at different points in time
after applying the voltage. The anchoring strengths at the substrates
are assumed to be equal, $W_L=W_U\equiv W$, and
the list of the parameters are given  in Table~\ref{tab:params}.
}
\label{fig:theta_vs_t}
\end{figure*}

\begin{figure*}[!thb]
\centering
\subfigure[]{%
\resizebox{85mm}{!}{\includegraphics*{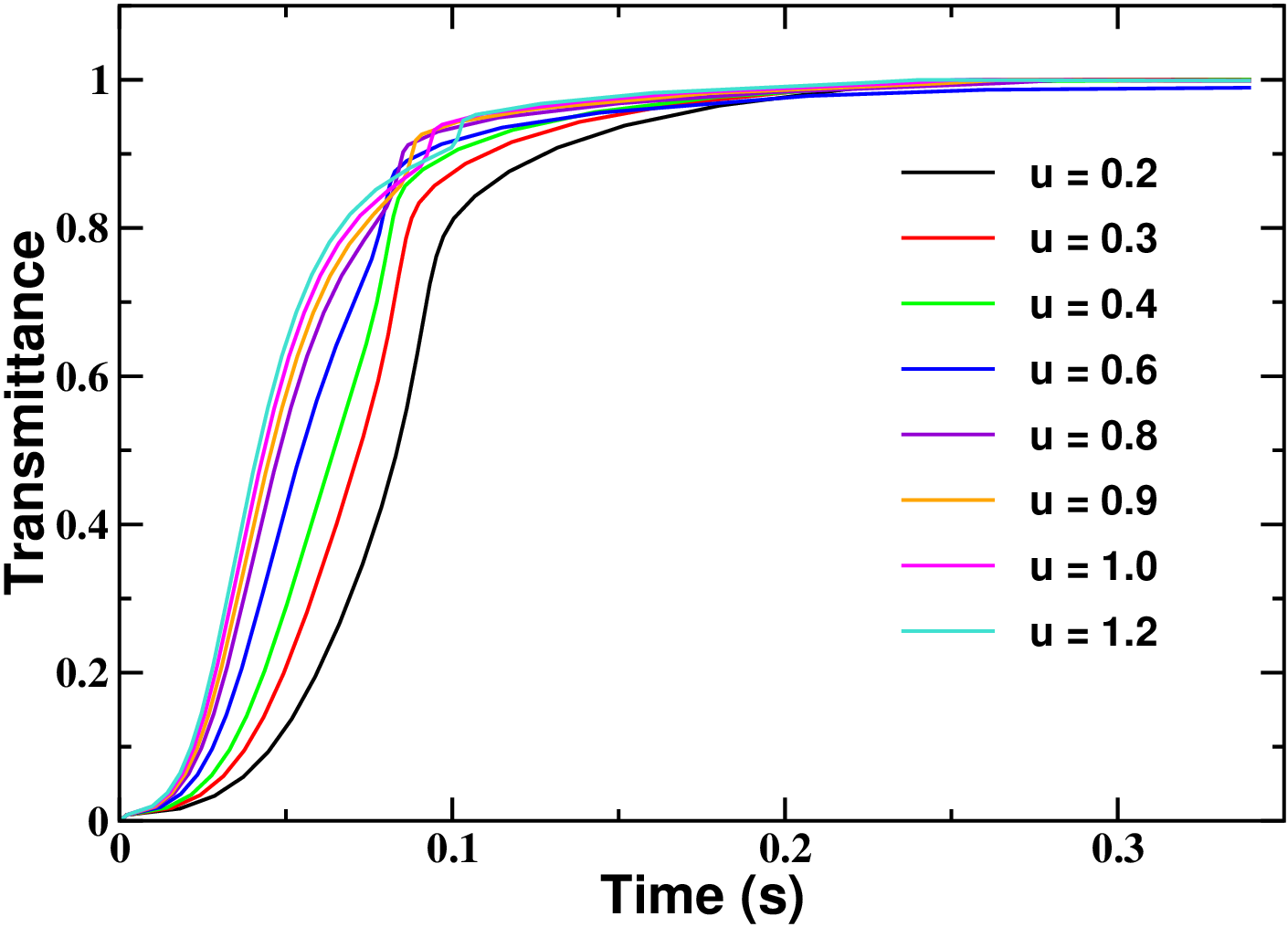}}
\label{fig:transm_u}}
\subfigure[]{%
\resizebox{85mm}{!}{\includegraphics*{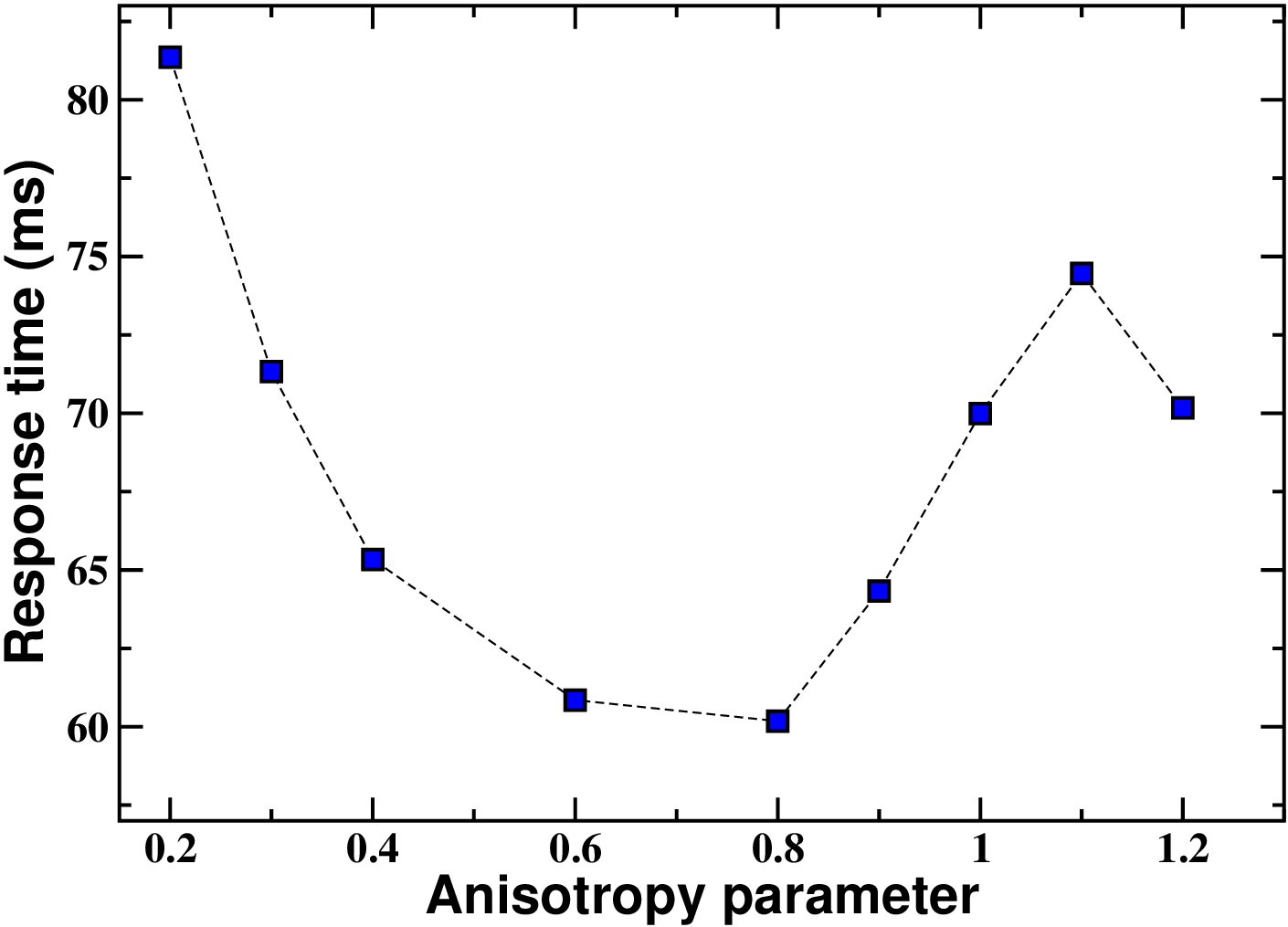}}
\label{fig:time_u}}
\caption{
(a) Transmittance as a function of time at
various values of the dielectric anisotropy parameter,
$u=(\epsilon_{\parallel}-\epsilon_{\perp})/\epsilon_{\perp}$.
(b) Response time as a function of $u$.
}
\label{fig:trm_vs_u}
\end{figure*}

\begin{figure*}[!thb]
\centering
\subfigure[]{%
\resizebox{85mm}{!}{\includegraphics*{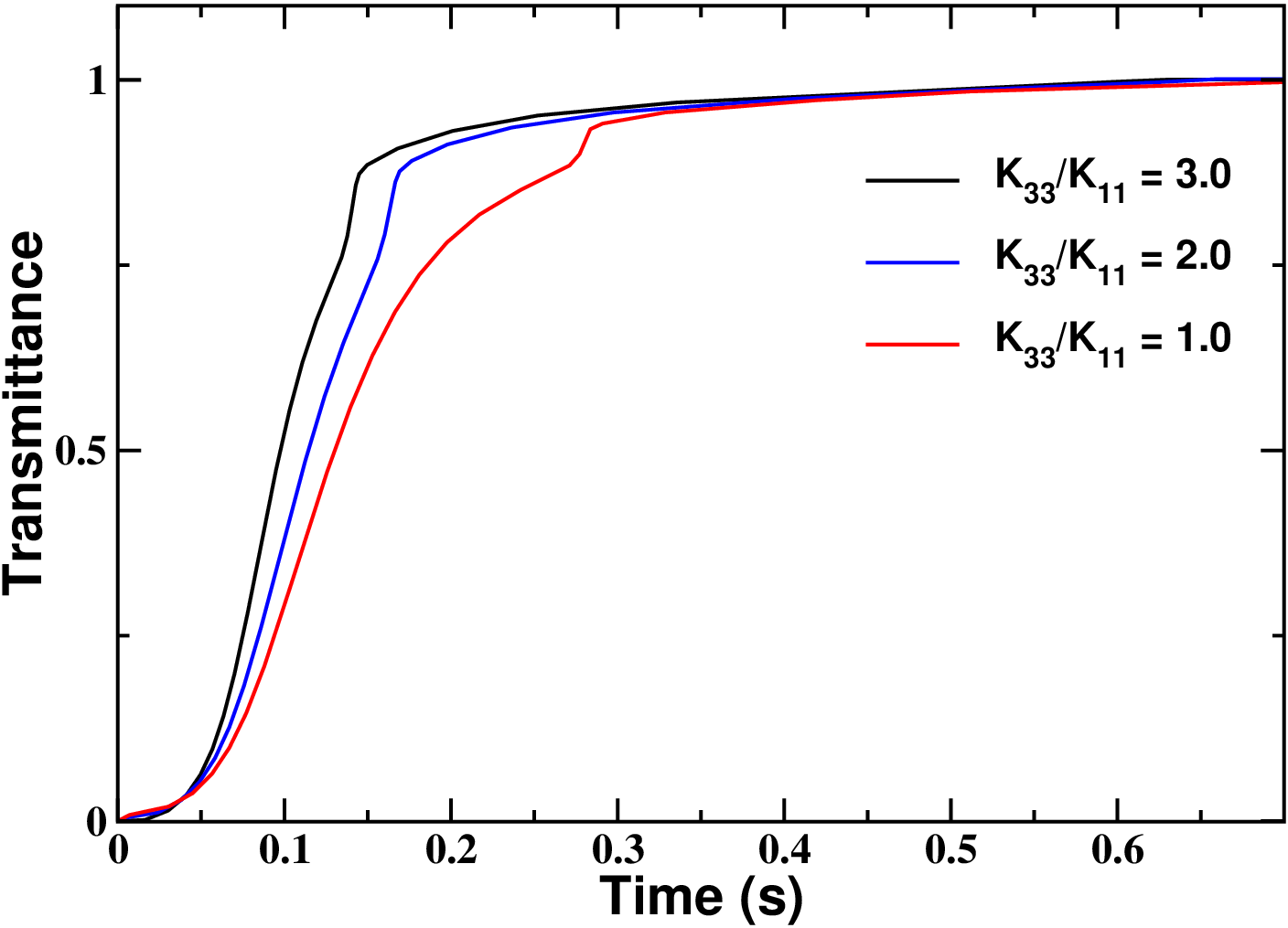}}
\label{fig:transm_k}}
\subfigure[]{%
\resizebox{85mm}{!}{\includegraphics*{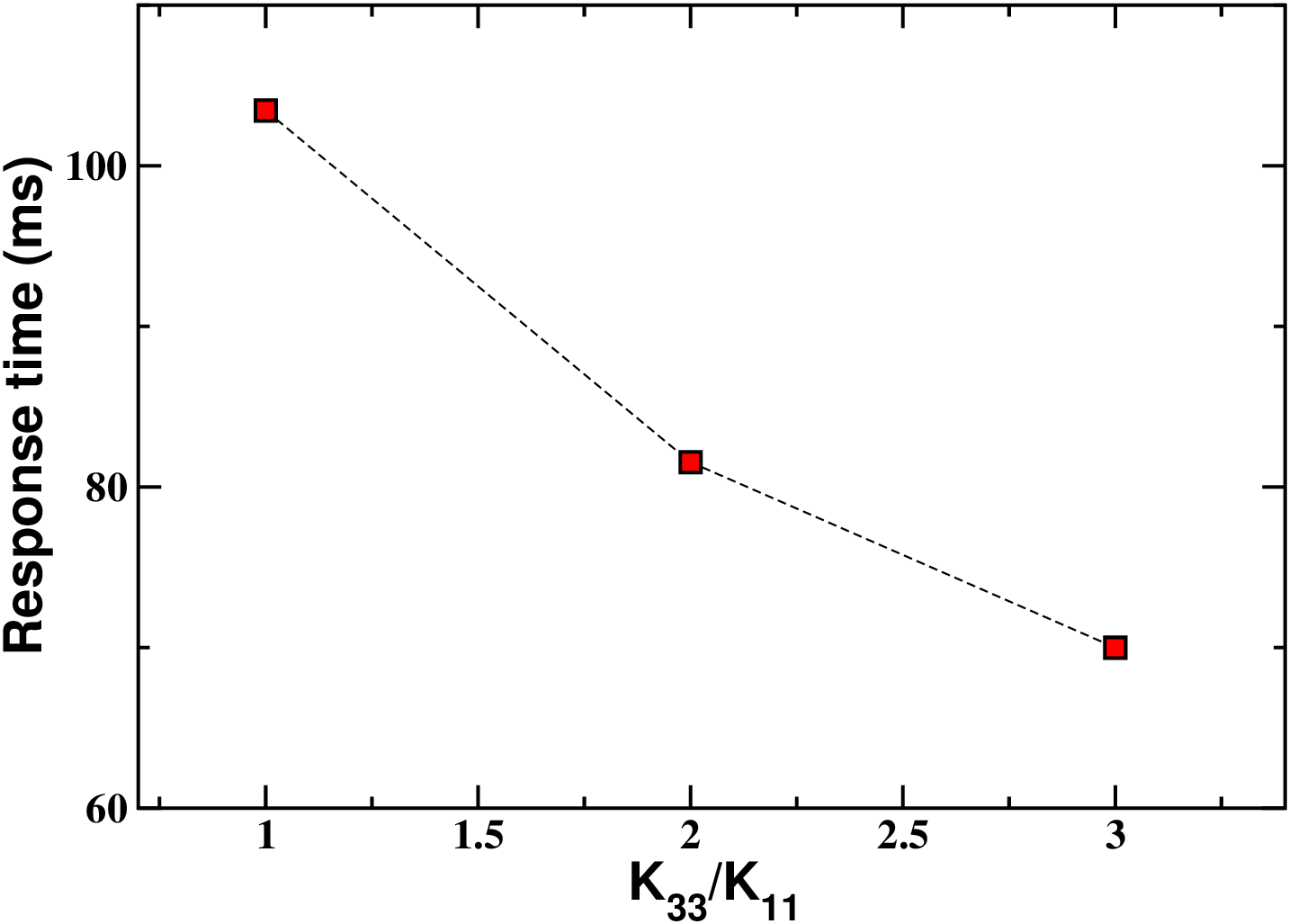}}
\label{fig:time_k}}
\caption{
(a) Transmittance as a function of time at
various values of the elastic ratio $K_{33}/K_{11}$,
($=(\theta_{L}-\theta_{U})$).
(b) Response time as a function of the elastic ratio $K_{33}/K_{11}$.
}
\label{fig:trm_vs_k}
\end{figure*}

\begin{figure*}[!thb]
\centering
\subfigure[]{%
\resizebox{85mm}{!}{\includegraphics*{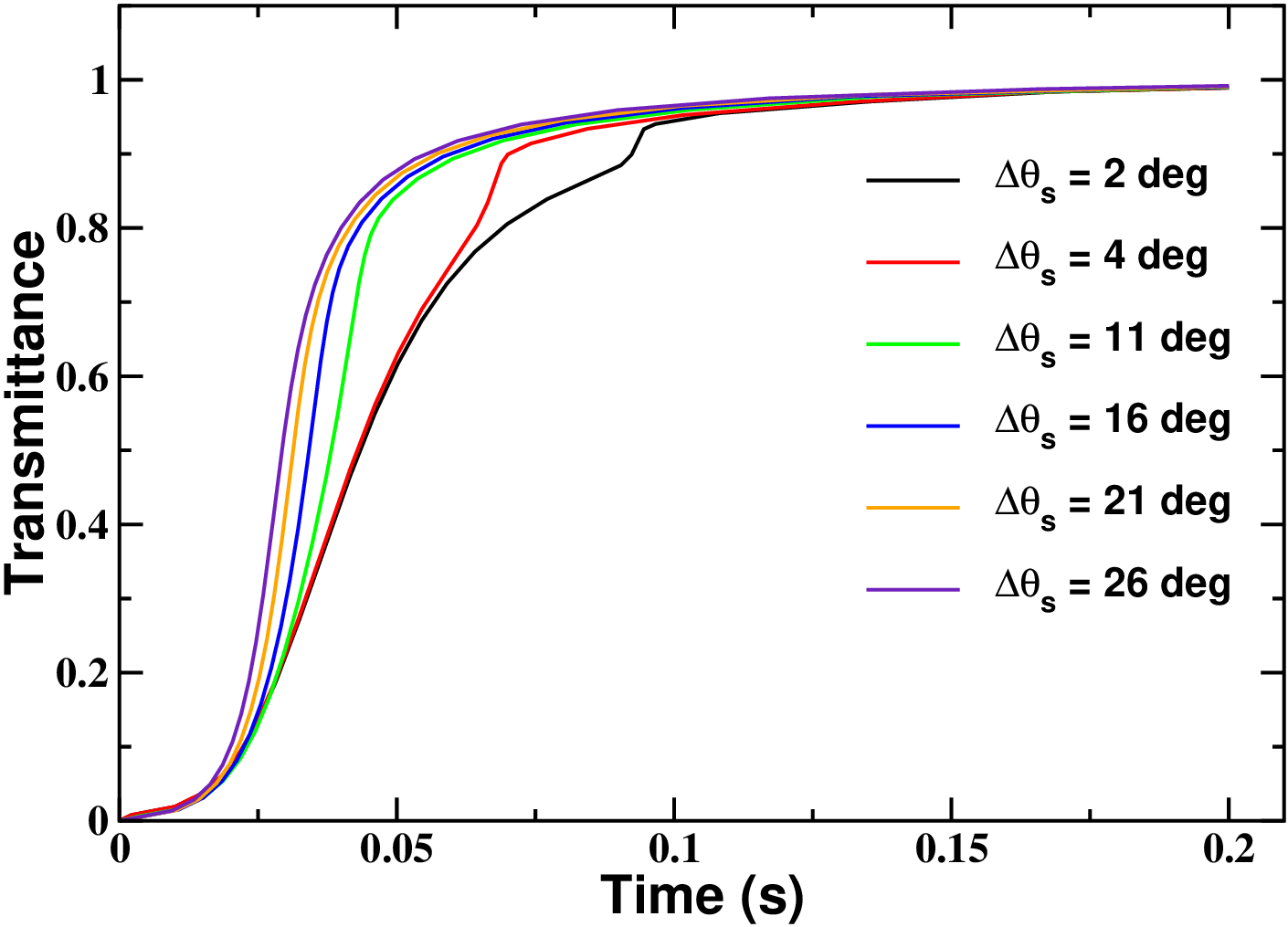}}
\label{fig:transm_dlt}}
\subfigure[]{%
\resizebox{85mm}{!}{\includegraphics*{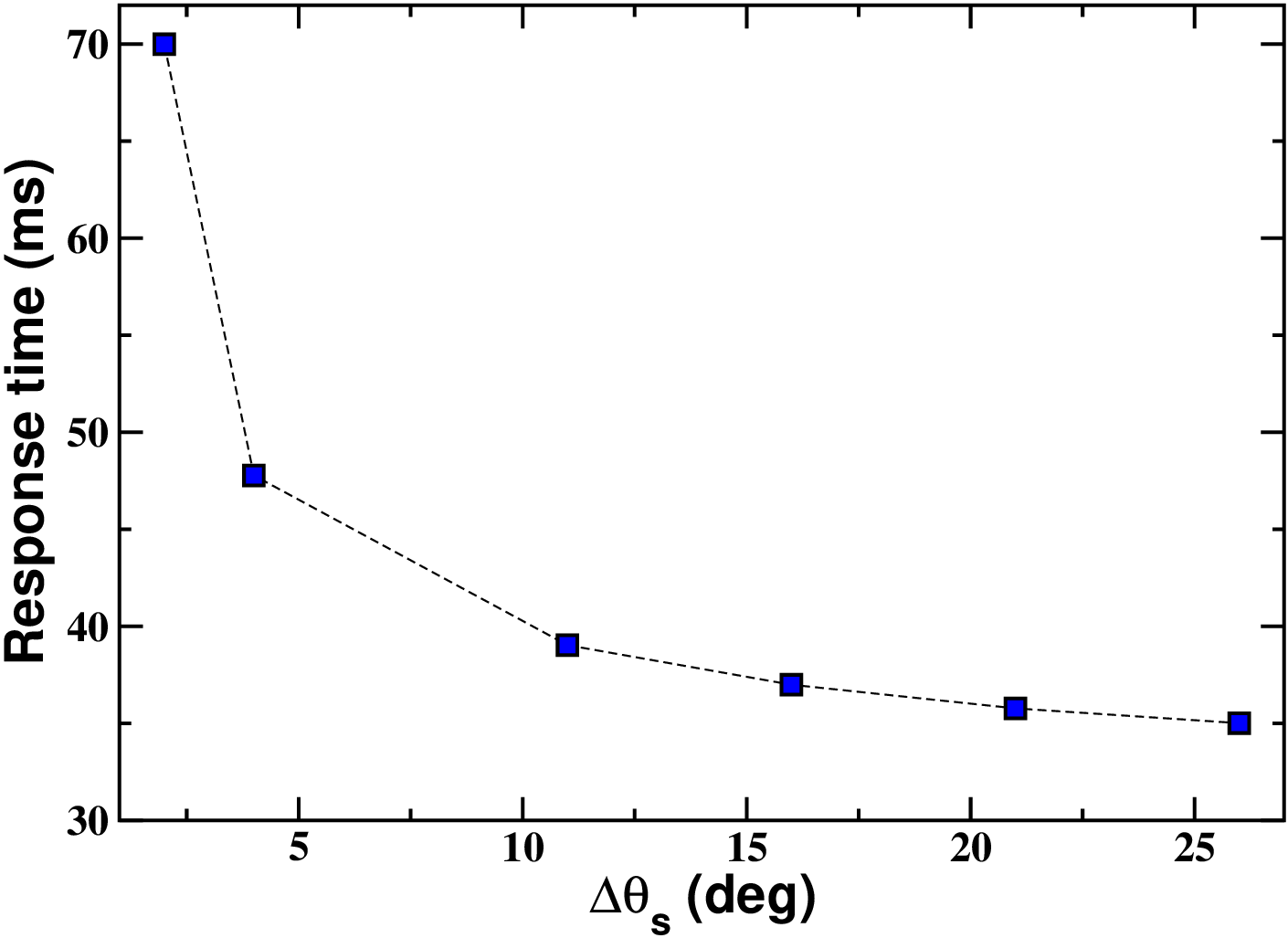}}
\label{fig:time_dlt}}
\caption{
(a) Transmittance as a function of time at
various values of $\Delta\theta_{s}$,
($=\theta_{L}-\theta_{U}$).
(b) Response time as a function of $\Delta\theta_{s}$.
}
\label{fig:trm_vs_deltathet}
\end{figure*}

\begin{figure*}[!tbh]
\centering
\resizebox{150mm}{!}{\includegraphics*{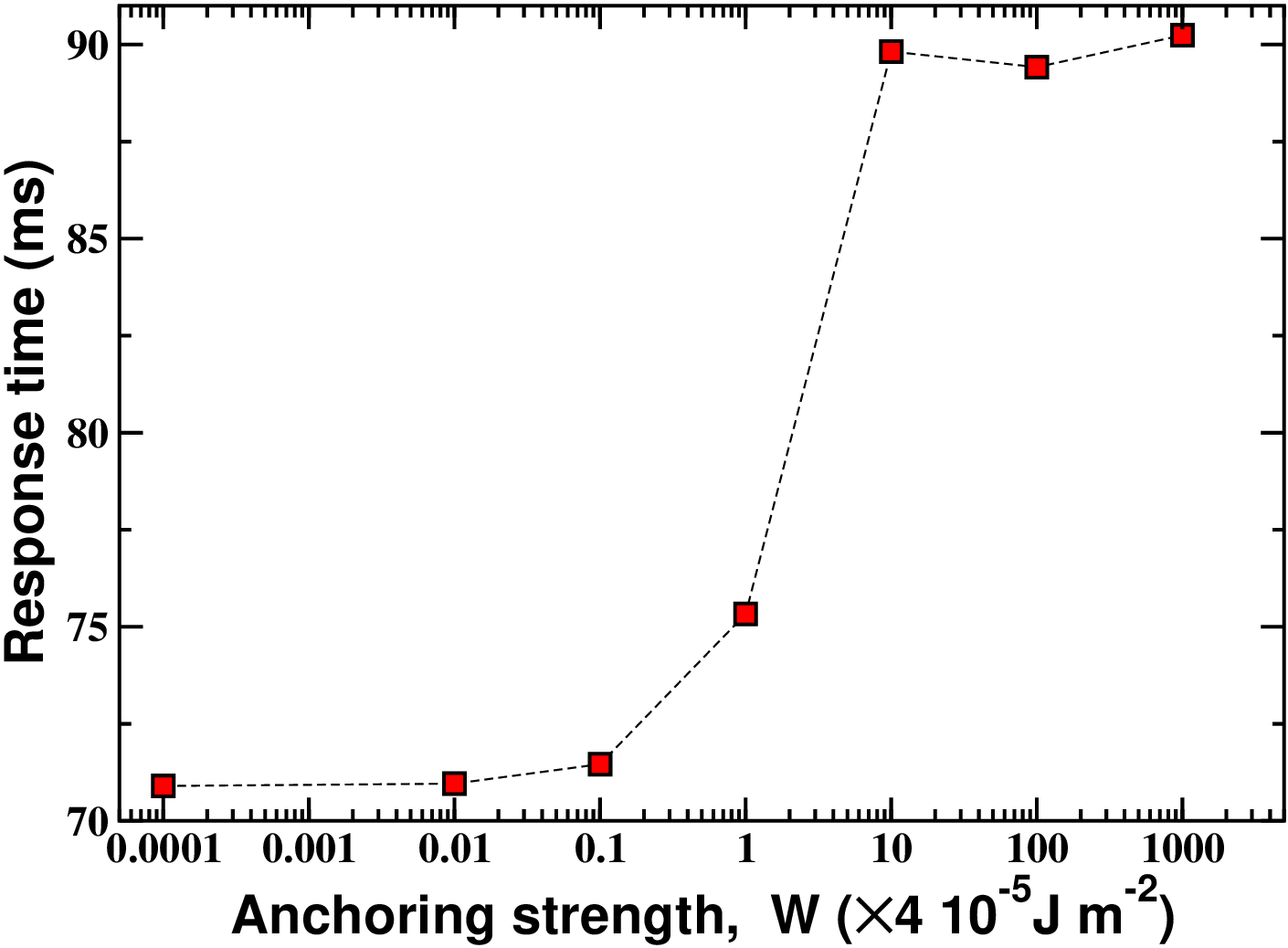}}
\caption{%
Response time as a function of
the anchoring strength, $W=W_L=W_U$.
}
\label{fig:time_w}
\end{figure*}

\begin{figure*}[!tbh]
\centering
\resizebox{150mm}{!}{\includegraphics*{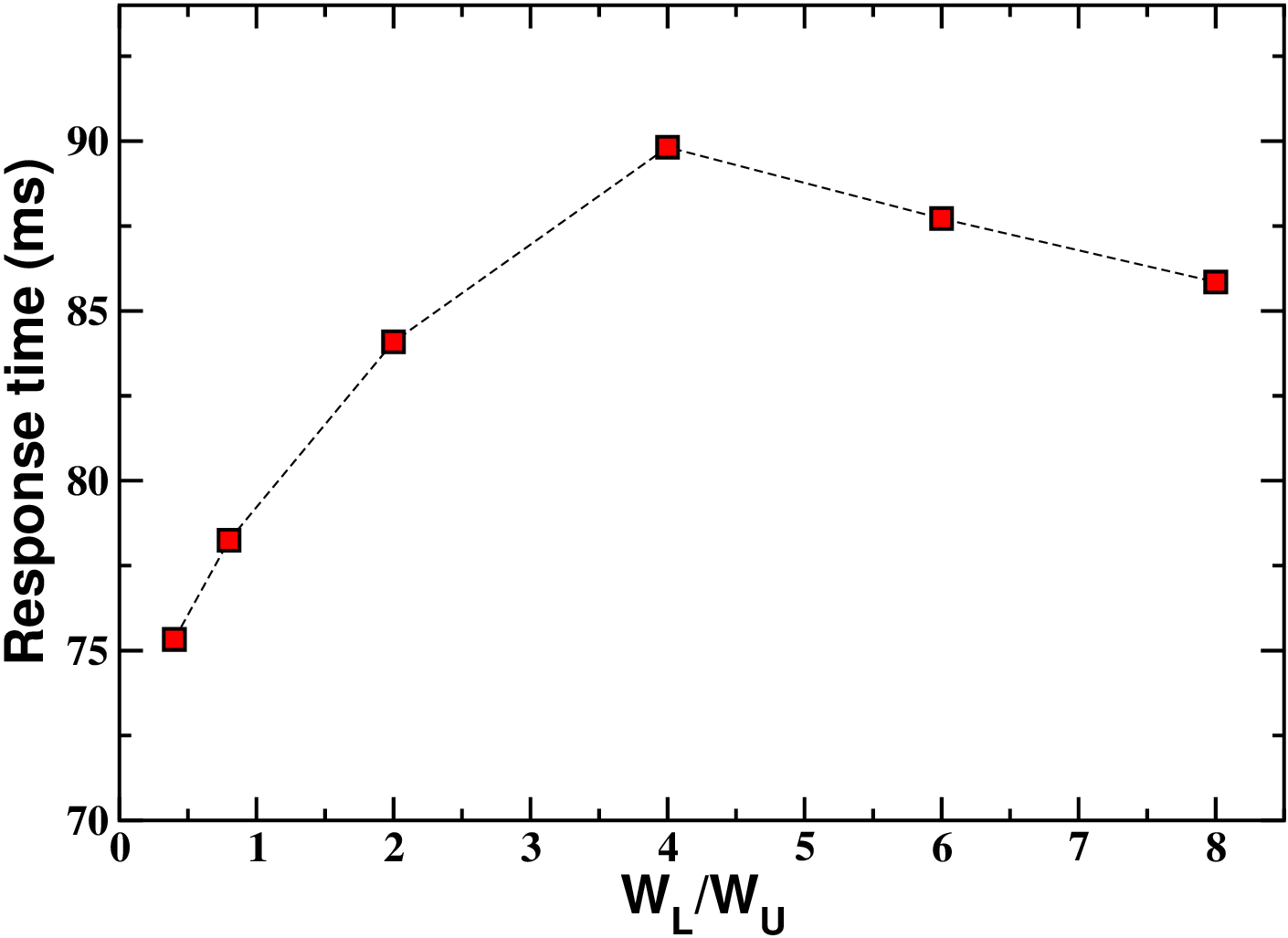}}
\caption{%
Response time as a function of
the anchoring strength ratio, $W_L/W_U$,
at $W_U=W=4.0\times10^{-4}$J/m$^2$.
}
\label{fig:time_wr}
\end{figure*}

\begin{figure*}[!tbh]
\centering
\resizebox{150mm}{!}{\includegraphics*{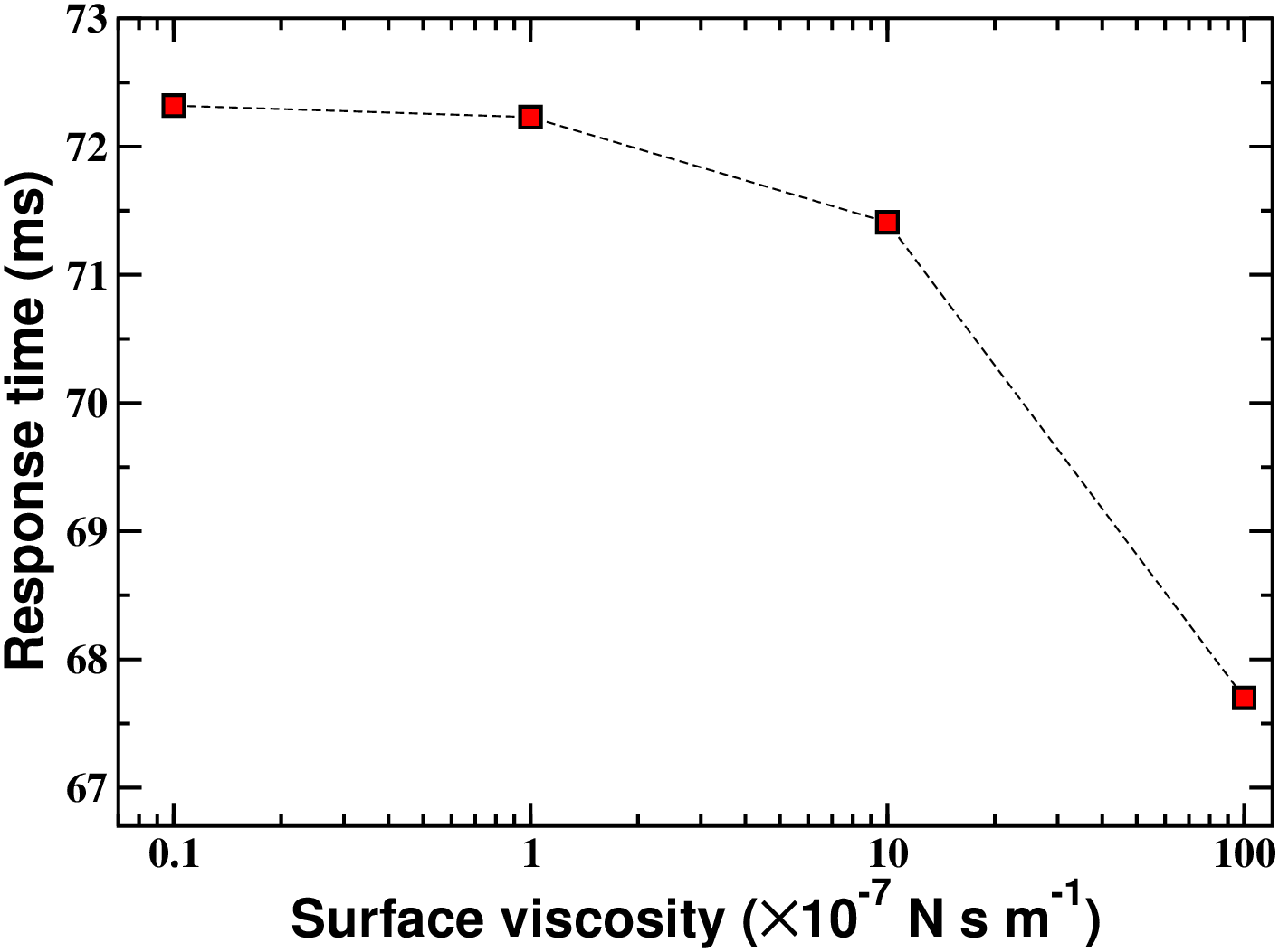}}
\caption{%
Response time as a function of
the rotational surface viscosity.
}
\label{fig:time_visc}
\end{figure*}

\begin{figure*}[!thb]
\centering
\subfigure[]{%
\resizebox{85mm}{!}{\includegraphics*{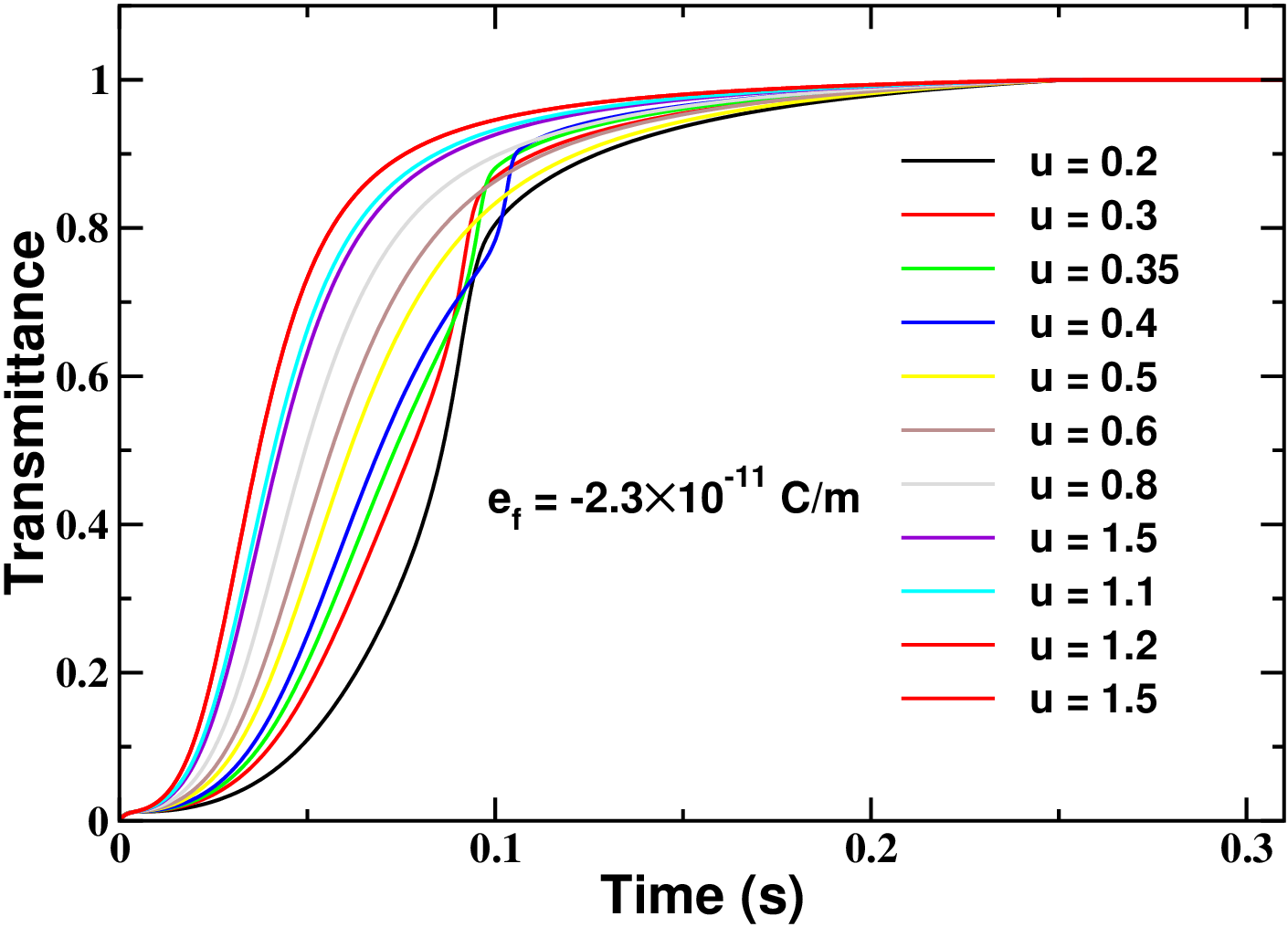}}
\label{fig:transm_u2}}
\subfigure[]{%
\resizebox{85mm}{!}{\includegraphics*{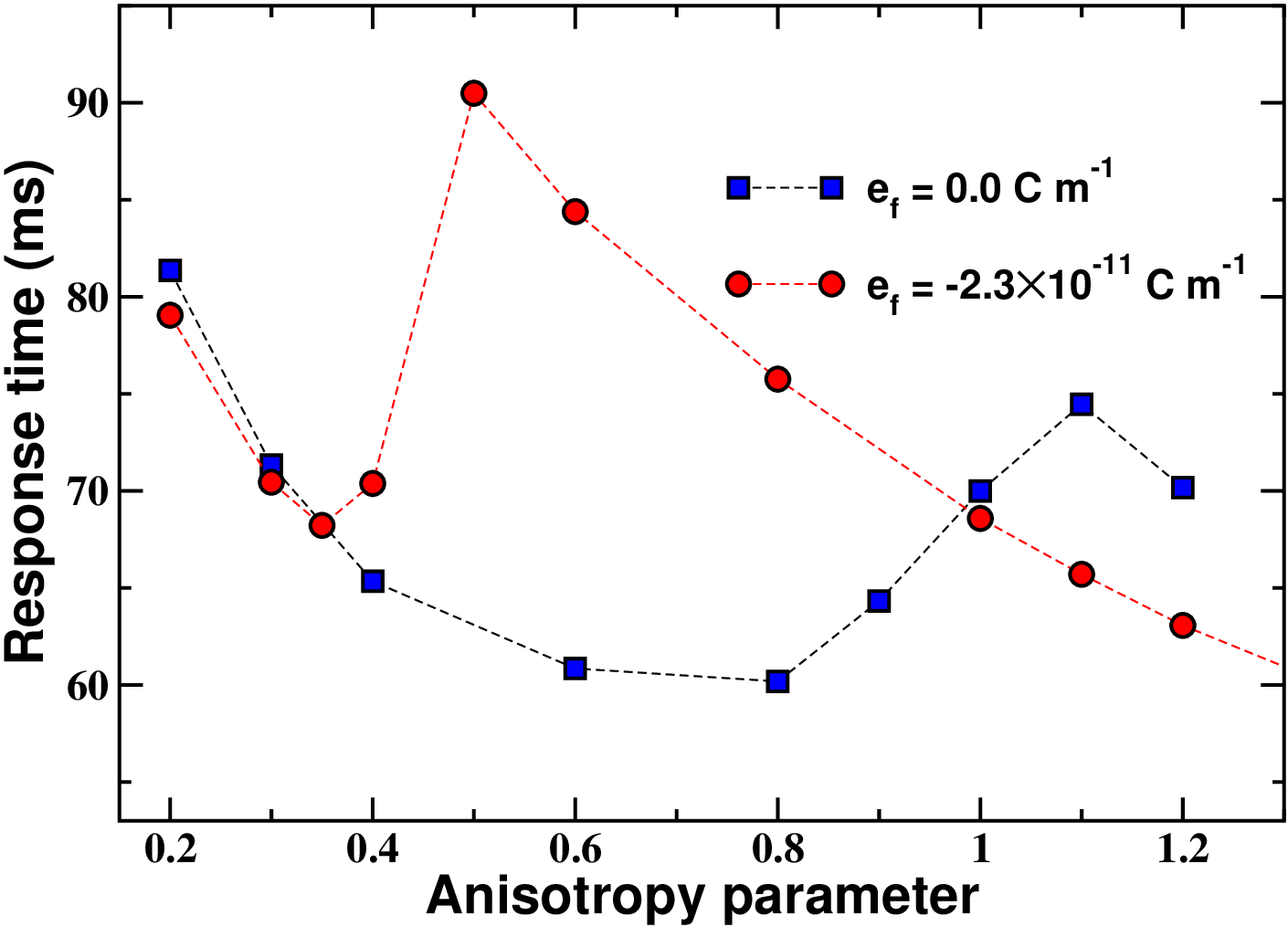}}
\label{fig:time_u2}}
\caption{
(a) Transmittance as a function of time for
various values of the dielectric anisotropy parameter
at non-vanishing flexoelectric coefficient $e_f$. 
(b) Response time as a function of $u$ at
$e_f=0.0$ C/m (squares) and $e_f=-2.3\times 10^{-12}$ C/m (circles).
}
\label{fig:trm_vs_u2}
\end{figure*}

\begin{figure*}[!tbh]
\centering
\resizebox{150mm}{!}{\includegraphics*{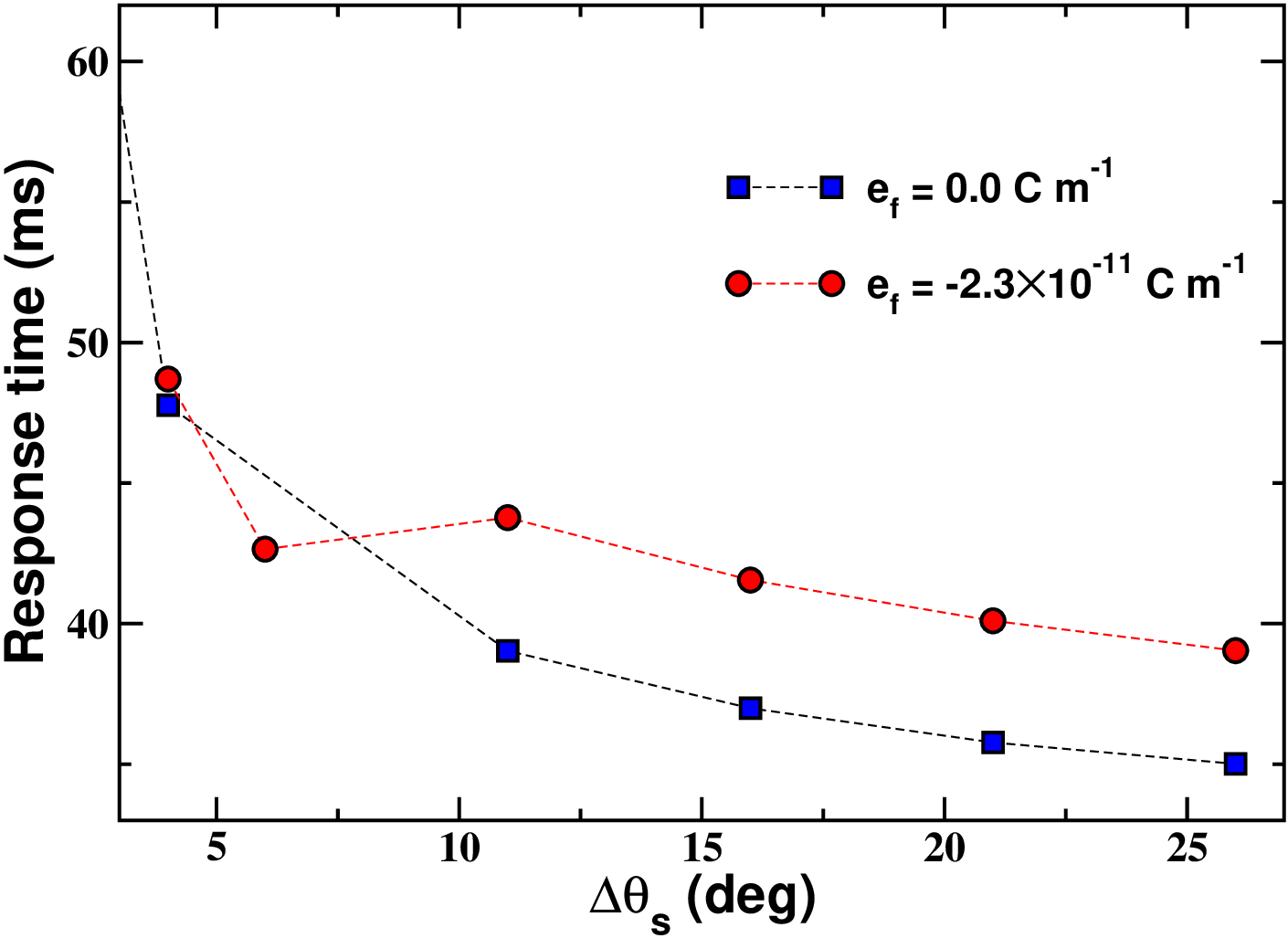}}
\caption{%
Response time as a function of
the pretilt angle difference at
$e_f=0.0$ C/m (squares) and $e_f=-2.3\times 10^{-12}$ C/m (circles).
}
\label{fig:time_dlt2}
\end{figure*}

\begin{figure*}[!thb]
\centering
\subfigure[]{%
\resizebox{85mm}{!}{\includegraphics*{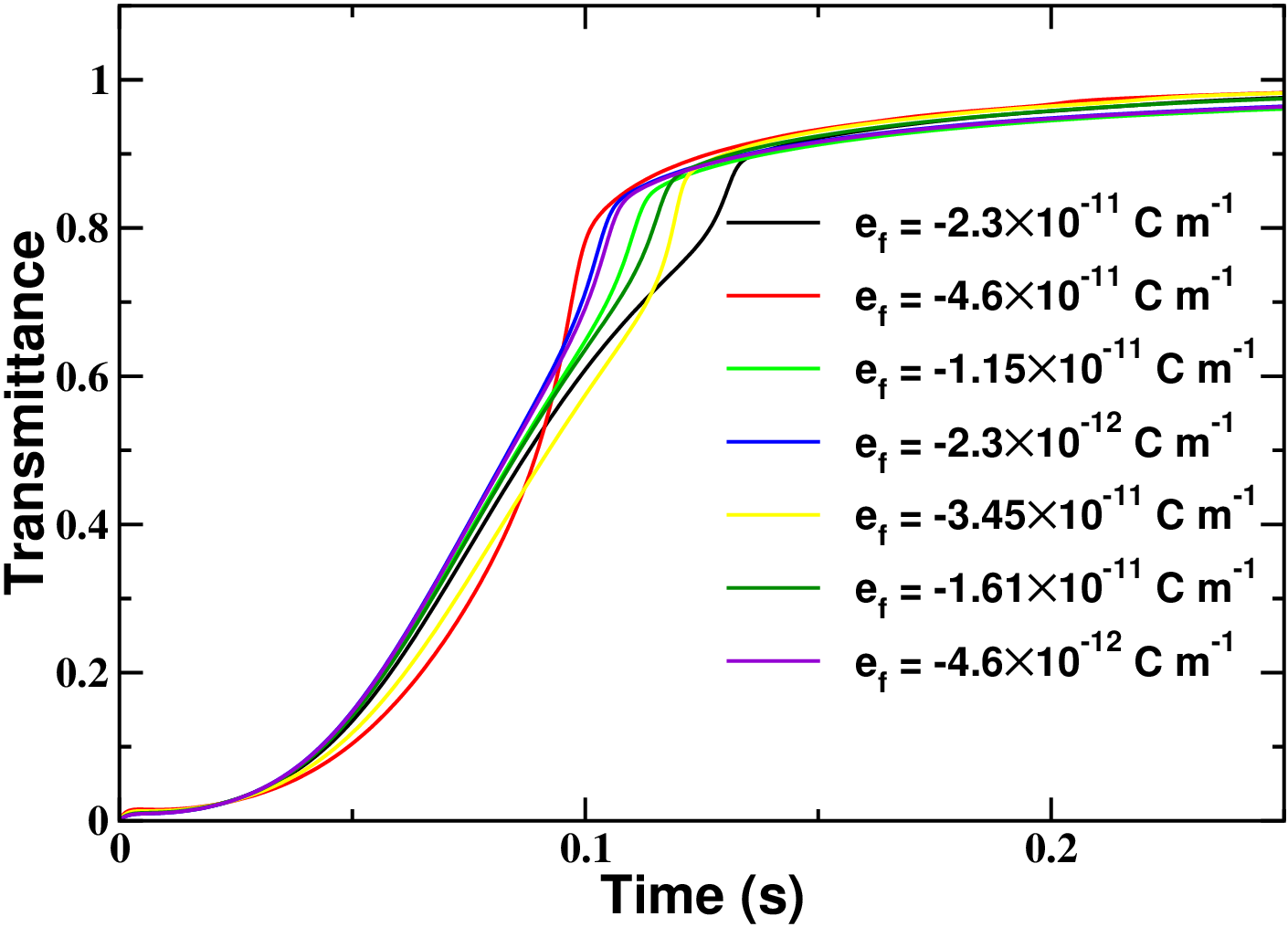}}
\label{fig:transm_ef}}
\subfigure[]{%
\resizebox{85mm}{!}{\includegraphics*{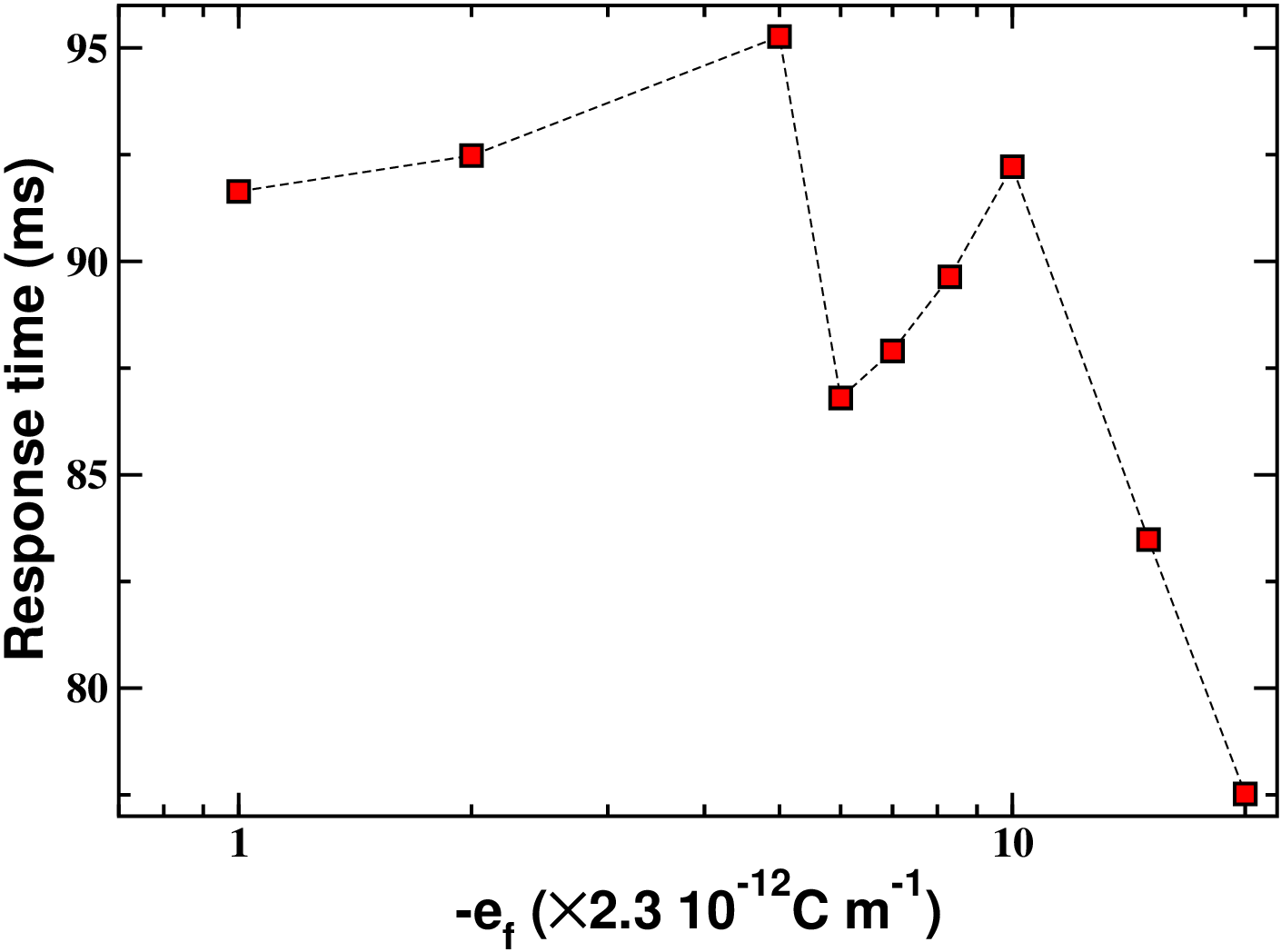}}
\label{fig:time_ef}}
\caption{
(a) Transmittance as a function of time at
various values of the flexoelectric coefficient $e_f$. 
(b) Response time as a function of the flexoelectric coefficient.
}
\label{fig:trm_vs_ef}
\end{figure*}

\section{Simulation results}
\label{sec:simulation-results}

In this section we present our numerical results obtained by
solving the dynamic equations~\eqref{eq:dyn_bulk}
and~\eqref{eq:dyn_surf} numerically. 
Dependencies of the tilt angle on $z$ at specified points in time
were computed  using the finite difference time domain method.
The parameters used in our calculations are listed in 
Table~\ref{tab:params}. 

In order to study the dynamics of optical response of the layer,
the data representing temporal evolution of the director profile,
which is the tilt angle as a function of $z$, $\theta(z,t)$,
were used as an input for computing 
the \textit{transmittance} of light through the layer placed
between two crossed polarisers.

The expression for the transmittance can be derived by using 
the Jones matrix method~\cite{Jones:1942}. 
When the director of LC cell is at $45$ degrees
to the input polariser, the transmittance, $T$,
is given by~\cite{Blin:b:1994,Kwok:jap:1996}
\begin{align}
&
  \label{eq:transmitt}
  T=\sin^2(\Delta\phi/2),
\\
&
\label{eq:path}
\Delta\phi=\frac{2\pi}{\lambda}\int_0^d (n_{\mrm{eff}}-n_{o})\dd z,
\\
&
\label{eq:n_eff}
\frac{1}{n^2_{\mrm{eff}}}=\frac{\sin^2\theta}{n^2_o}+
\frac{\cos^2\theta}{n^2_e},
\end{align}
where $\Delta\phi$ is the phase difference between the ordinary and
extraordinary ray; $\lambda$ is the wavelength of the incident light 
and $n_o$ ($n_e$) is the ordinary (extraordinary) refractive index.

Thus, we describe the dynamics of optical response
by computing temporal change in the transmittance~\eqref{eq:transmitt}. 
An important parameter characterising the rate of change of 
the transmittance is the \textit{response time} which is the time it takes for the
transmittance to increase from 10\% to 90\%.

We begin with the case in which the flexoelectric effect is neglected
and $e_f=0$.  
Fig.~\ref{fig:theta_vs_t} shows how the director profile evolves in time after
applying the voltage across the NLC cell. 
Asymmetric
pretilt angles and symmetric surface anchoring energy are used in this
calculation. 
As is illustrated in Fig.~\ref{fig:theta_vs_t}, the initial director
configuration corresponds to the splay state
which gradually transforms into the bend state under the
action of the electric field. 

Now we pass on to discussing the effects related to
the dielectric and elastic anisotropies.
The results for various values of the dielectric anisotropy
parameter, $u=(\epsilon_{\parallel}-\epsilon_{\perp})/\epsilon_{\perp}$, 
and the elastic ratio, $K_{33}/K_{11}$,
are shown in Figs.~\ref{fig:trm_vs_u} and~\ref{fig:trm_vs_k},
respectively.

Fig.~\ref{fig:time_u} indicates that the response time
is a non-monotonic function of the dielectric anisotropy parameter
and goes through a minimum in the vicinity of $u=0.8$.
By contrast, as is shown in Fig.~\ref{fig:time_k}, the response time
monotonically declines as the ratio of $K_{33}$ and $K_{11}$ increases.
So, large values of the elastic ratio facilitate the splay-bend
transition.


The surface pretilt angles, $\theta_L$ and $\theta_U$,
are known to play an important part in the splay-bend 
transition~\cite{Acosta:lc:2000}.
These are among the parameters that affect the dynamics of optical
response through the boundary conditions at the
substrates~\eqref{eq:dyn_surf}.

The first parameter we consider is the difference between
the pretilt angles: $\Delta\theta_s=\theta_L - \theta_U$.  
Fig.~\ref{fig:transm_dlt} shows the curves for
the transmittance varying in time at various values
of the pretilt angle difference.
As is evident from Fig.~\ref{fig:transm_dlt}, 
the curves are getting steeper as  $\Delta\theta_s$ increases
and the response time, shown in Fig.~\ref{fig:time_dlt}, 
is a decreasing function of $\Delta\theta_s$.

The anchoring energy dependence of the response time is plotted in
Fig.~\ref{fig:time_w} for the symmetric case with $W_L=W_U\equiv W$.
The curve is depicted in logarithmic scale and 
clearly indicates the transition between two regimes of anchoring:
the weak anchoring regime and the strong anchoring regime.
In the regime of weak anchoring, the extrapolation length is
larger than the cell thickness, $d$, and the response time is small.
As is seen from Fig.~\ref{fig:time_w},
the response time increases with the anchoring energy
and saturates on reaching the strong anchoring regime 
where the extrapolation length is much smaller
than $d$.

Influence of asymmetry in the anchoring energy strengths
on the response time is illustrated in Fig.~\ref{fig:time_wr}
where the anchoring strength at the upper substrate is kept constant
at the value listed in Table~\ref{tab:params}, $W_U=W$.
It is shown that
the response time varies slowly and 
reaches its maximum at $W_L/W_U\approx 4.0$.

The surface rotational viscosity, $\gamma_s$,
can be conveniently characterised by the ratio of $\gamma_s$ and the
bulk viscosity which has the dimension of length.
There are, however, only few measurements of
this length that, according
to~\cite{Pert:lc:1995,Mert:2000,Cop:pre:2001},
can be of the order tens and hundreds nanometers.
Our numerical results on the surface viscosity dependence of
the response time are presented in Fig.~\ref{fig:time_visc}.
It is seen that  variations of the surface viscosity 
over a wide range of values
have almost no effect on the response time. 

So far we have limited our discussion
to the case in which the flexoelectric coefficient $e_f$
vanishes and thus the flexoelectric effect
appears to be eliminated from  the consideration. 
There are some measurements of the flexoelectric 
coefficient in a variety of liquid 
crystals~\cite{Madh:jpp:1985,Madh:jdp:1997,Taka:jjap:1998,Blin:pre:2001,Sambl:pre:2001,Sambl:jap:2002} .
It was found that  the value of $|e_f|$  typically falls in the range between $5\times10^{-12}$ C/m and
$9\times10^{-11}$ C/m.
But reliable and accurate experimental estimates of $e_f$
are still missing. 
For example,
the reported values of $e_f$ for MBBA turned out  to 
differ in both magnitude and sign depending
on theoretical approach used for processing 
experimental data~\cite{Madh:jpp:1985,Taka:jjap:1998,Zum:lc:1999,Kirk:lc:2003}.

Numerical results related to the effect of flexoelectricity
on the dynamics of NLC cell are presented in 
Figs.~\ref{fig:trm_vs_u2}--\ref{fig:trm_vs_ef}.
The curves shown in Fig.~\ref{fig:trm_vs_u2}
indicate that the dielectric anisotropy dependence
of the response time turns out to be strongly affected
by the flexoelectric effect.
In the presence of flexoelectricity the curve has a pronounced maximum
peaked at $u\approx 0.5$ which follows a minimum reached at 
$u\approx 0.35$.

By contrast to the dielectric anisotropy dependence,
the dependencies of the response time on the
pretilt angle difference, $\Delta\theta_s$,
depicted in Fig.~\ref{fig:time_dlt2}, do not differ
significantly. For $\Delta\theta_s$ larger than 10 deg, 
referring to Fig.~\ref{fig:time_dlt2},
the curve with non-zero flexoelectric coefficient
is approximately shifted upward by
$5$ ms with respect to the curve computed at 
vanishing $e_f$.

Finally, we comment on the dependencies displayed
in Fig.~\ref{fig:trm_vs_ef}.
The curves plotted in Fig.~\ref{fig:transm_ef} represent
temporal evolution of the transmittance at
different values of the flexoelectric coefficient.
The response time in relation to the flexoelectric coefficient
obtained from these curves is shown in Fig.~\ref{fig:time_ef}.
It can be seen that the response time steeply declines
after reaching a maximum at $e_f\approx -1.15\times 10^{-11}$C/m.

\section{Conclusion}
\label{sec:discussion}

In this paper we used a simplified approach to study the dynamics of optical response
at the splay-bend transition that occurs after applying the voltage 
across the NLC cell. It is assumed that the coupling between the
director and the flow velocity can be eliminated from consideration.
 
Similar approach was recently applied to
formulate the model of 
switching in a zenithally bistable device~\cite{Mottr:pre:2002}.
In our case, however, not only the boundary conditions~\eqref{eq:dyn_surf} 
are different, but also inhomogeneity of the electric field
is taken into account using the constitutive relation~\eqref{eq:constit_rel}.

The simulation results for the transmittance 
were obtained by solving the dynamic equations of the model
numerically.  The response time characterising the rate of change of
the transmittance was evaluated to study  how the parameters of the
cell influence the  dynamics of optical response.

Dependencies of the response time on the dielectric anisotropy parameter
and on the flexoelectric coefficient are found to be strongly non-monotonic.
It was shown that the response time  declines as
the elastic ratio $K_{33}/K_{11}$ or the pretilt angle difference $\Delta\theta_s$ increases. 
From the other hand, the response time appears to be relatively insensitive 
to anchoring strength asymmetry and to changes in the surface viscosity. 


\begin{acknowledgments}
This research was partially supported by RGC Grants
HKUST6004/01E and HKUST6102/03E.
\end{acknowledgments}


\begin{thebibliography}{48}
\expandafter\ifx\csname natexlab\endcsname\relax\def\natexlab#1{#1}\fi
\expandafter\ifx\csname bibnamefont\endcsname\relax
  \def\bibnamefont#1{#1}\fi
\expandafter\ifx\csname bibfnamefont\endcsname\relax
  \def\bibfnamefont#1{#1}\fi
\expandafter\ifx\csname citenamefont\endcsname\relax
  \def\citenamefont#1{#1}\fi
\expandafter\ifx\csname url\endcsname\relax
  \def\url#1{\texttt{#1}}\fi
\expandafter\ifx\csname urlprefix\endcsname\relax\def\urlprefix{URL }\fi
\providecommand{\bibinfo}[2]{#2}
\providecommand{\eprint}[2][]{\url{#2}}

\bibitem[{\citenamefont{Berreman and Heffner}(1981)}]{Berrem:jap:1981}
\bibinfo{author}{\bibfnamefont{D.~W.} \bibnamefont{Berreman}} \bibnamefont{and}
  \bibinfo{author}{\bibfnamefont{W.~R.} \bibnamefont{Heffner}},
  \bibinfo{journal}{J. Appl. Phys.} \textbf{\bibinfo{volume}{52}},
  \bibinfo{pages}{3032} (\bibinfo{year}{1981}).

\bibitem[{\citenamefont{Xie and Kwok}(1998)}]{Xie:jap:1998}
\bibinfo{author}{\bibfnamefont{Z.~L.} \bibnamefont{Xie}} \bibnamefont{and}
  \bibinfo{author}{\bibfnamefont{H.~S.} \bibnamefont{Kwok}},
  \bibinfo{journal}{J. Appl. Phys.} \textbf{\bibinfo{volume}{84}},
  \bibinfo{pages}{77} (\bibinfo{year}{1998}).

\bibitem[{\citenamefont{Zhuang et~al.}(1999)\citenamefont{Zhuang, Kim, and
  Patel}}]{Zhuang:apl:1999}
\bibinfo{author}{\bibfnamefont{Z.}~\bibnamefont{Zhuang}},
  \bibinfo{author}{\bibfnamefont{Y.~J.} \bibnamefont{Kim}}, \bibnamefont{and}
  \bibinfo{author}{\bibfnamefont{J.~S.} \bibnamefont{Patel}},
  \bibinfo{journal}{Appl. Phys. Lett.} \textbf{\bibinfo{volume}{75}},
  \bibinfo{pages}{3008} (\bibinfo{year}{1999}).

\bibitem[{\citenamefont{Xie et~al.}(2000{\natexlab{a}})\citenamefont{Xie, Dong,
  Xu, Gao, and Kwok}}]{Xie:jap2:2000}
\bibinfo{author}{\bibfnamefont{Z.~L.} \bibnamefont{Xie}},
  \bibinfo{author}{\bibfnamefont{Y.~M.} \bibnamefont{Dong}},
  \bibinfo{author}{\bibfnamefont{S.~Y.} \bibnamefont{Xu}},
  \bibinfo{author}{\bibfnamefont{H.~J.} \bibnamefont{Gao}}, \bibnamefont{and}
  \bibinfo{author}{\bibfnamefont{H.~S.} \bibnamefont{Kwok}},
  \bibinfo{journal}{J. Appl. Phys.} \textbf{\bibinfo{volume}{87}},
  \bibinfo{pages}{2673} (\bibinfo{year}{2000}{\natexlab{a}}).

\bibitem[{\citenamefont{Xie et~al.}(2000{\natexlab{b}})\citenamefont{Xie,
  Zheng, Xu, Gao, and Kwok}}]{Xie:jap1:2000}
\bibinfo{author}{\bibfnamefont{Z.~L.} \bibnamefont{Xie}},
  \bibinfo{author}{\bibfnamefont{C.~Y.} \bibnamefont{Zheng}},
  \bibinfo{author}{\bibfnamefont{S.~Y.} \bibnamefont{Xu}},
  \bibinfo{author}{\bibfnamefont{H.~J.} \bibnamefont{Gao}}, \bibnamefont{and}
  \bibinfo{author}{\bibfnamefont{H.~S.} \bibnamefont{Kwok}},
  \bibinfo{journal}{J. Appl. Phys.} \textbf{\bibinfo{volume}{88}},
  \bibinfo{pages}{1722} (\bibinfo{year}{2000}{\natexlab{b}}).

\bibitem[{\citenamefont{Guo et~al.}(2000)\citenamefont{Guo, Meng, Wong, and
  Kwok}}]{Guo:apl:2000}
\bibinfo{author}{\bibfnamefont{J.~X.} \bibnamefont{Guo}},
  \bibinfo{author}{\bibfnamefont{Z.~G.} \bibnamefont{Meng}},
  \bibinfo{author}{\bibfnamefont{M.}~\bibnamefont{Wong}}, \bibnamefont{and}
  \bibinfo{author}{\bibfnamefont{H.~S.} \bibnamefont{Kwok}},
  \bibinfo{journal}{Appl. Phys. Lett.} \textbf{\bibinfo{volume}{77}},
  \bibinfo{pages}{3716} (\bibinfo{year}{2000}).

\bibitem[{\citenamefont{Cheng et~al.}(1981)\citenamefont{Cheng, Thurston, and
  Berreman}}]{Cheng:jap:1981}
\bibinfo{author}{\bibfnamefont{J.}~\bibnamefont{Cheng}},
  \bibinfo{author}{\bibfnamefont{R.~N.} \bibnamefont{Thurston}},
  \bibnamefont{and} \bibinfo{author}{\bibfnamefont{D.~W.}
  \bibnamefont{Berreman}}, \bibinfo{journal}{J. Appl. Phys.}
  \textbf{\bibinfo{volume}{52}}, \bibinfo{pages}{2756} (\bibinfo{year}{1981}).

\bibitem[{\citenamefont{Acosta et~al.}(2000)\citenamefont{Acosta, Towler, and
  Walton}}]{Acosta:lc:2000}
\bibinfo{author}{\bibfnamefont{E.~J.} \bibnamefont{Acosta}},
  \bibinfo{author}{\bibfnamefont{M.~J.} \bibnamefont{Towler}},
  \bibnamefont{and} \bibinfo{author}{\bibfnamefont{H.~G.}
  \bibnamefont{Walton}}, \bibinfo{journal}{Liq. Cryst.}
  \textbf{\bibinfo{volume}{27}}, \bibinfo{pages}{977} (\bibinfo{year}{2000}).

\bibitem[{\citenamefont{Nakamura and Noguchi}(2000)}]{Nakam:jjap:2000}
\bibinfo{author}{\bibfnamefont{H.}~\bibnamefont{Nakamura}} \bibnamefont{and}
  \bibinfo{author}{\bibfnamefont{M.}~\bibnamefont{Noguchi}},
  \bibinfo{journal}{Jpn. J. Appl. Phys.} \textbf{\bibinfo{volume}{39}},
  \bibinfo{pages}{6368} (\bibinfo{year}{2000}).

\bibitem[{\citenamefont{Lee et~al.}(2003)\citenamefont{Lee, Kim, Lee, Yoon, and
  Kim}}]{Lee:jjap:2003}
\bibinfo{author}{\bibfnamefont{S.~H.} \bibnamefont{Lee}},
  \bibinfo{author}{\bibfnamefont{T.~J.} \bibnamefont{Kim}},
  \bibinfo{author}{\bibfnamefont{G.~D.} \bibnamefont{Lee}},
  \bibinfo{author}{\bibfnamefont{T.~H.} \bibnamefont{Yoon}}, \bibnamefont{and}
  \bibinfo{author}{\bibfnamefont{J.~C.} \bibnamefont{Kim}},
  \bibinfo{journal}{Jpn. J. Appl. Phys.} \textbf{\bibinfo{volume}{42}},
  \bibinfo{pages}{L1148} (\bibinfo{year}{2003}).

\bibitem[{\citenamefont{Inoue et~al.}(2003)\citenamefont{Inoue, Miyashita,
  Uchida, Yamada, and Ishii}}]{Inou:sid:2003}
\bibinfo{author}{\bibfnamefont{I.}~\bibnamefont{Inoue}},
  \bibinfo{author}{\bibfnamefont{T.}~\bibnamefont{Miyashita}},
  \bibinfo{author}{\bibfnamefont{T.}~\bibnamefont{Uchida}},
  \bibinfo{author}{\bibfnamefont{Y.}~\bibnamefont{Yamada}}, \bibnamefont{and}
  \bibinfo{author}{\bibfnamefont{Y.}~\bibnamefont{Ishii}},
  \bibinfo{journal}{Journal of the SID} \textbf{\bibinfo{volume}{11/3}},
  \bibinfo{pages}{571} (\bibinfo{year}{2003}).

\bibitem[{\citenamefont{Porte and Jadot}(1978)}]{Port:jpp:1978}
\bibinfo{author}{\bibfnamefont{G.}~\bibnamefont{Porte}} \bibnamefont{and}
  \bibinfo{author}{\bibfnamefont{J.~P.} \bibnamefont{Jadot}},
  \bibinfo{journal}{J. Phys. (Paris)} \textbf{\bibinfo{volume}{39}},
  \bibinfo{pages}{213} (\bibinfo{year}{1978}).

\bibitem[{\citenamefont{Komitov et~al.}(1986)\citenamefont{Komitov, Hauck, and
  Koswig}}]{Komit:pss:1986}
\bibinfo{author}{\bibfnamefont{L.}~\bibnamefont{Komitov}},
  \bibinfo{author}{\bibfnamefont{G.}~\bibnamefont{Hauck}}, \bibnamefont{and}
  \bibinfo{author}{\bibfnamefont{H.~D.} \bibnamefont{Koswig}},
  \bibinfo{journal}{Phys. Stat. Sol.} \textbf{\bibinfo{volume}{97}},
  \bibinfo{pages}{645} (\bibinfo{year}{1986}).

\bibitem[{\citenamefont{Cheng and Gao}(2001)}]{Cheng:lc:2001}
\bibinfo{author}{\bibfnamefont{H.}~\bibnamefont{Cheng}} \bibnamefont{and}
  \bibinfo{author}{\bibfnamefont{H.}~\bibnamefont{Gao}}, \bibinfo{journal}{Liq.
  Cryst.} \textbf{\bibinfo{volume}{28}}, \bibinfo{pages}{1337}
  (\bibinfo{year}{2001}).

\bibitem[{\citenamefont{Davidson and Mottram}(2002)}]{Mottr:pre:2002}
\bibinfo{author}{\bibfnamefont{A.~J.} \bibnamefont{Davidson}} \bibnamefont{and}
  \bibinfo{author}{\bibfnamefont{N.~J.} \bibnamefont{Mottram}},
  \bibinfo{journal}{Phys. Rev. E} \textbf{\bibinfo{volume}{65}},
  \bibinfo{pages}{051710} (\bibinfo{year}{2002}).

\bibitem[{\citenamefont{Cheng and Gao}(2003)}]{Cheng:lc:2003}
\bibinfo{author}{\bibfnamefont{H.}~\bibnamefont{Cheng}} \bibnamefont{and}
  \bibinfo{author}{\bibfnamefont{H.}~\bibnamefont{Gao}}, \bibinfo{journal}{Liq.
  Cryst.} \textbf{\bibinfo{volume}{30}}, \bibinfo{pages}{839}
  (\bibinfo{year}{2003}).

\bibitem[{\citenamefont{Tsoy}(2002)}]{Tsoy:2002}
\bibinfo{author}{\bibfnamefont{V.~I.} \bibnamefont{Tsoy}},
  \bibinfo{journal}{Techn. Phys.} \textbf{\bibinfo{volume}{47}},
  \bibinfo{pages}{34} (\bibinfo{year}{2002}).

\bibitem[{\citenamefont{Rapini and Papoular}(1969)}]{Rap:1969}
\bibinfo{author}{\bibfnamefont{A.}~\bibnamefont{Rapini}} \bibnamefont{and}
  \bibinfo{author}{\bibfnamefont{M.}~\bibnamefont{Papoular}},
  \bibinfo{journal}{J. Phys. (Paris) Colloq. C4} \textbf{\bibinfo{volume}{30}},
  \bibinfo{pages}{54} (\bibinfo{year}{1969}).

\bibitem[{\citenamefont{de~Gennes and Prost}(1993)}]{Gennes:bk:1993}
\bibinfo{author}{\bibfnamefont{P.~G.} \bibnamefont{de~Gennes}}
  \bibnamefont{and} \bibinfo{author}{\bibfnamefont{J.}~\bibnamefont{Prost}},
  \emph{\bibinfo{title}{The Physics of Liquid Crystals}}
  (\bibinfo{publisher}{Clarendon Press}, \bibinfo{address}{Oxford},
  \bibinfo{year}{1993}).

\bibitem[{\citenamefont{Meyer}(1969)}]{Meyer:prl:1969}
\bibinfo{author}{\bibfnamefont{R.~B.} \bibnamefont{Meyer}},
  \bibinfo{journal}{Phys. Rev. Lett.} \textbf{\bibinfo{volume}{22}},
  \bibinfo{pages}{918} (\bibinfo{year}{1969}).

\bibitem[{\citenamefont{Kirkman et~al.}(2003)\citenamefont{Kirkman, Stirner,
  and Hagston}}]{Kirk:lc:2003}
\bibinfo{author}{\bibfnamefont{N.~T.} \bibnamefont{Kirkman}},
  \bibinfo{author}{\bibfnamefont{T.}~\bibnamefont{Stirner}}, \bibnamefont{and}
  \bibinfo{author}{\bibfnamefont{W.~E.} \bibnamefont{Hagston}},
  \bibinfo{journal}{Liq. Cryst.} \textbf{\bibinfo{volume}{30}},
  \bibinfo{pages}{1115} (\bibinfo{year}{2003}).

\bibitem[{\citenamefont{Thurston and Berreman}(1981)}]{Thur:jap:1981}
\bibinfo{author}{\bibfnamefont{R.~N.} \bibnamefont{Thurston}} \bibnamefont{and}
  \bibinfo{author}{\bibfnamefont{D.~W.} \bibnamefont{Berreman}},
  \bibinfo{journal}{J. Appl. Phys.} \textbf{\bibinfo{volume}{52}},
  \bibinfo{pages}{508} (\bibinfo{year}{1981}).

\bibitem[{\citenamefont{Palierne}(1986)}]{Palier:prl:1986}
\bibinfo{author}{\bibfnamefont{J.~F.} \bibnamefont{Palierne}},
  \bibinfo{journal}{Phys. Rev. Lett.} \textbf{\bibinfo{volume}{56}},
  \bibinfo{pages}{1160} (\bibinfo{year}{1986}).

\bibitem[{\citenamefont{Barbero and Durand}(1987)}]{Dur:pra:1987}
\bibinfo{author}{\bibfnamefont{G.}~\bibnamefont{Barbero}} \bibnamefont{and}
  \bibinfo{author}{\bibfnamefont{G.}~\bibnamefont{Durand}},
  \bibinfo{journal}{Phys. Rev. A} \textbf{\bibinfo{volume}{35}},
  \bibinfo{pages}{1294} (\bibinfo{year}{1987}).

\bibitem[{\citenamefont{Barbero and Durand}(1990)}]{Dur:jap:1990}
\bibinfo{author}{\bibfnamefont{G.}~\bibnamefont{Barbero}} \bibnamefont{and}
  \bibinfo{author}{\bibfnamefont{G.}~\bibnamefont{Durand}},
  \bibinfo{journal}{J. Appl. Phys.} \textbf{\bibinfo{volume}{68}},
  \bibinfo{pages}{5549} (\bibinfo{year}{1990}).

\bibitem[{\citenamefont{Ponti et~al.}(1999)\citenamefont{Ponti, Ziherl,
  Ferrero, and \v{Z}umer}}]{Zum:lc:1999}
\bibinfo{author}{\bibfnamefont{S.}~\bibnamefont{Ponti}},
  \bibinfo{author}{\bibfnamefont{P.}~\bibnamefont{Ziherl}},
  \bibinfo{author}{\bibfnamefont{C.}~\bibnamefont{Ferrero}}, \bibnamefont{and}
  \bibinfo{author}{\bibfnamefont{S.}~\bibnamefont{\v{Z}umer}},
  \bibinfo{journal}{Liq. Cryst.} \textbf{\bibinfo{volume}{26}},
  \bibinfo{pages}{1171} (\bibinfo{year}{1999}).

\bibitem[{\citenamefont{Brown and Mottram}(2003)}]{Mottr:pre:2003}
\bibinfo{author}{\bibfnamefont{C.~V.} \bibnamefont{Brown}} \bibnamefont{and}
  \bibinfo{author}{\bibfnamefont{N.~J.} \bibnamefont{Mottram}},
  \bibinfo{journal}{Phys. Rev. E} \textbf{\bibinfo{volume}{68}},
  \bibinfo{pages}{031702} (\bibinfo{year}{2003}).

\bibitem[{\citenamefont{Felczak and Derfel}(2003)}]{Felc:lc:2003}
\bibinfo{author}{\bibfnamefont{M.}~\bibnamefont{Felczak}} \bibnamefont{and}
  \bibinfo{author}{\bibfnamefont{G.}~\bibnamefont{Derfel}},
  \bibinfo{journal}{Liq. Cryst.} \textbf{\bibinfo{volume}{30}},
  \bibinfo{pages}{739} (\bibinfo{year}{2003}).

\bibitem[{\citenamefont{Hohenberg and Halperin}(1977)}]{Hohen:rmp:1977}
\bibinfo{author}{\bibfnamefont{P.~C.} \bibnamefont{Hohenberg}}
  \bibnamefont{and} \bibinfo{author}{\bibfnamefont{B.~I.}
  \bibnamefont{Halperin}}, \bibinfo{journal}{Rev. Mod. Phys}
  \textbf{\bibinfo{volume}{49}}, \bibinfo{pages}{435} (\bibinfo{year}{1977}).

\bibitem[{\citenamefont{Chaikin and Lubensky}(1995)}]{Luben:bk:1995}
\bibinfo{author}{\bibfnamefont{P.~M.} \bibnamefont{Chaikin}} \bibnamefont{and}
  \bibinfo{author}{\bibfnamefont{T.~C.} \bibnamefont{Lubensky}},
  \emph{\bibinfo{title}{Principles of Condensed Matter Physics}}
  (\bibinfo{publisher}{Cambridge University Press},
  \bibinfo{address}{Cambridge}, \bibinfo{year}{1995}).

\bibitem[{\citenamefont{Pieransky et~al.}(1973)\citenamefont{Pieransky,
  Brochard, and Guyon}}]{Pier:jpp:1973}
\bibinfo{author}{\bibfnamefont{P.}~\bibnamefont{Pieransky}},
  \bibinfo{author}{\bibfnamefont{F.}~\bibnamefont{Brochard}}, \bibnamefont{and}
  \bibinfo{author}{\bibfnamefont{E.}~\bibnamefont{Guyon}}, \bibinfo{journal}{J.
  Phys. (Paris)} \textbf{\bibinfo{volume}{34}}, \bibinfo{pages}{35}
  (\bibinfo{year}{1973}).

\bibitem[{\citenamefont{Berreman}(1975)}]{Berrem:jap:1975}
\bibinfo{author}{\bibfnamefont{D.~W.} \bibnamefont{Berreman}},
  \bibinfo{journal}{J. Appl. Phys.} \textbf{\bibinfo{volume}{46}},
  \bibinfo{pages}{3746} (\bibinfo{year}{1975}).

\bibitem[{\citenamefont{Chen and Hsieh}(1991)}]{Chen:pra:1991}
\bibinfo{author}{\bibfnamefont{S.~M.} \bibnamefont{Chen}} \bibnamefont{and}
  \bibinfo{author}{\bibfnamefont{T.~C.} \bibnamefont{Hsieh}},
  \bibinfo{journal}{Phys. Rev. A} \textbf{\bibinfo{volume}{43}},
  \bibinfo{pages}{2848} (\bibinfo{year}{1991}).

\bibitem[{\citenamefont{Mertelj and \v{C}opi\v{c}}(2000)}]{Mert:2000}
\bibinfo{author}{\bibfnamefont{A.}~\bibnamefont{Mertelj}} \bibnamefont{and}
  \bibinfo{author}{\bibfnamefont{M.}~\bibnamefont{\v{C}opi\v{c}}},
  \bibinfo{journal}{Phys. Rev. E} \textbf{\bibinfo{volume}{61}},
  \bibinfo{pages}{1622} (\bibinfo{year}{2000}).

\bibitem[{\citenamefont{Vilfan et~al.}(2001)\citenamefont{Vilfan, Olenik,
  Mertelj, and \v{C}opi\v{c}}}]{Cop:pre:2001}
\bibinfo{author}{\bibfnamefont{M.}~\bibnamefont{Vilfan}},
  \bibinfo{author}{\bibfnamefont{I.~D.} \bibnamefont{Olenik}},
  \bibinfo{author}{\bibfnamefont{A.}~\bibnamefont{Mertelj}}, \bibnamefont{and}
  \bibinfo{author}{\bibfnamefont{M.}~\bibnamefont{\v{C}opi\v{c}}},
  \bibinfo{journal}{Phys. Rev. E} \textbf{\bibinfo{volume}{63}},
  \bibinfo{pages}{061709} (\bibinfo{year}{2001}).

\bibitem[{\citenamefont{Ziherl and \v{Z}umer}(1996)}]{Zih:pre:1996}
\bibinfo{author}{\bibfnamefont{P.}~\bibnamefont{Ziherl}} \bibnamefont{and}
  \bibinfo{author}{\bibfnamefont{S.}~\bibnamefont{\v{Z}umer}},
  \bibinfo{journal}{Phys. Rev. E} \textbf{\bibinfo{volume}{54}},
  \bibinfo{pages}{1592} (\bibinfo{year}{1996}).

\bibitem[{\citenamefont{Sonnet et~al.}(2000)\citenamefont{Sonnet, Virga, and
  Durand}}]{Dur:pre:2000}
\bibinfo{author}{\bibfnamefont{A.~M.} \bibnamefont{Sonnet}},
  \bibinfo{author}{\bibfnamefont{E.~G.} \bibnamefont{Virga}}, \bibnamefont{and}
  \bibinfo{author}{\bibfnamefont{G.}~\bibnamefont{Durand}},
  \bibinfo{journal}{Phys. Rev. E} \textbf{\bibinfo{volume}{62}},
  \bibinfo{pages}{3694} (\bibinfo{year}{2000}).

\bibitem[{\citenamefont{Durand and Virga}(1999)}]{Dur:pre:1999}
\bibinfo{author}{\bibfnamefont{G.}~\bibnamefont{Durand}} \bibnamefont{and}
  \bibinfo{author}{\bibfnamefont{E.~G.} \bibnamefont{Virga}},
  \bibinfo{journal}{Phys. Rev. E} \textbf{\bibinfo{volume}{59}},
  \bibinfo{pages}{4137} (\bibinfo{year}{1999}).

\bibitem[{\citenamefont{Jones}(1942)}]{Jones:1942}
\bibinfo{author}{\bibfnamefont{C.~R.} \bibnamefont{Jones}},
  \bibinfo{journal}{J. Opt. Soc. Am.} \textbf{\bibinfo{volume}{32}},
  \bibinfo{pages}{486} (\bibinfo{year}{1942}).

\bibitem[{\citenamefont{Blinov and Chigrinov}(1994)}]{Blin:b:1994}
\bibinfo{author}{\bibfnamefont{L.~M.} \bibnamefont{Blinov}} \bibnamefont{and}
  \bibinfo{author}{\bibfnamefont{V.~G.} \bibnamefont{Chigrinov}},
  \emph{\bibinfo{title}{Electrooptic effects in liquid crystal materials}}
  (\bibinfo{publisher}{Springer-Verlag}, \bibinfo{address}{Berlin},
  \bibinfo{year}{1994}).

\bibitem[{\citenamefont{Kwok}(1996)}]{Kwok:jap:1996}
\bibinfo{author}{\bibfnamefont{H.~S.} \bibnamefont{Kwok}}, \bibinfo{journal}{J.
  Appl. Phys.} \textbf{\bibinfo{volume}{80}}, \bibinfo{pages}{3687}
  (\bibinfo{year}{1996}).

\bibitem[{\citenamefont{Petrov et~al.}(1995)\citenamefont{Petrov, Ionescu,
  Versace, and Scaramuzza}}]{Pert:lc:1995}
\bibinfo{author}{\bibfnamefont{A.~G.} \bibnamefont{Petrov}},
  \bibinfo{author}{\bibfnamefont{A.~T.} \bibnamefont{Ionescu}},
  \bibinfo{author}{\bibfnamefont{C.}~\bibnamefont{Versace}}, \bibnamefont{and}
  \bibinfo{author}{\bibfnamefont{M.}~\bibnamefont{Scaramuzza}},
  \bibinfo{journal}{Liq. Cryst.} \textbf{\bibinfo{volume}{19}},
  \bibinfo{pages}{169} (\bibinfo{year}{1995}).

\bibitem[{\citenamefont{Madhusudana and Durand}(1985)}]{Madh:jpp:1985}
\bibinfo{author}{\bibfnamefont{N.~V.} \bibnamefont{Madhusudana}}
  \bibnamefont{and} \bibinfo{author}{\bibfnamefont{G.}~\bibnamefont{Durand}},
  \bibinfo{journal}{J. Phys. (Paris) Lett.} \textbf{\bibinfo{volume}{46}},
  \bibinfo{pages}{L195} (\bibinfo{year}{1985}).

\bibitem[{\citenamefont{Warrier and Madhusudana}(1997)}]{Madh:jdp:1997}
\bibinfo{author}{\bibfnamefont{S.~R.} \bibnamefont{Warrier}} \bibnamefont{and}
  \bibinfo{author}{\bibfnamefont{N.~V.} \bibnamefont{Madhusudana}},
  \bibinfo{journal}{J. Phys. II} \textbf{\bibinfo{volume}{7}},
  \bibinfo{pages}{1789} (\bibinfo{year}{1997}).

\bibitem[{\citenamefont{Takahashi et~al.}(1998)\citenamefont{Takahashi,
  Hashidate, Nishijou, Usui, Kimura, and Akahane}}]{Taka:jjap:1998}
\bibinfo{author}{\bibfnamefont{T.}~\bibnamefont{Takahashi}},
  \bibinfo{author}{\bibfnamefont{S.}~\bibnamefont{Hashidate}},
  \bibinfo{author}{\bibfnamefont{H.}~\bibnamefont{Nishijou}},
  \bibinfo{author}{\bibfnamefont{M.}~\bibnamefont{Usui}},
  \bibinfo{author}{\bibfnamefont{M.}~\bibnamefont{Kimura}}, \bibnamefont{and}
  \bibinfo{author}{\bibfnamefont{T.}~\bibnamefont{Akahane}},
  \bibinfo{journal}{Jpn. J. Appl. Phys.} \textbf{\bibinfo{volume}{37}},
  \bibinfo{pages}{1865} (\bibinfo{year}{1998}).

\bibitem[{\citenamefont{Blinov et~al.}(2001)\citenamefont{Blinov, Barnik,
  Ohoka, Ozaki, and Yoshino}}]{Blin:pre:2001}
\bibinfo{author}{\bibfnamefont{L.~M.} \bibnamefont{Blinov}},
  \bibinfo{author}{\bibfnamefont{M.~I.} \bibnamefont{Barnik}},
  \bibinfo{author}{\bibfnamefont{H.}~\bibnamefont{Ohoka}},
  \bibinfo{author}{\bibfnamefont{M.}~\bibnamefont{Ozaki}}, \bibnamefont{and}
  \bibinfo{author}{\bibfnamefont{K.}~\bibnamefont{Yoshino}},
  \bibinfo{journal}{Phys. Rev. E} \textbf{\bibinfo{volume}{64}},
  \bibinfo{pages}{031707} (\bibinfo{year}{2001}).

\bibitem[{\citenamefont{Mazzulla et~al.}(2001)\citenamefont{Mazzulla, Ciuchi,
  and Sambles}}]{Sambl:pre:2001}
\bibinfo{author}{\bibfnamefont{A.}~\bibnamefont{Mazzulla}},
  \bibinfo{author}{\bibfnamefont{F.}~\bibnamefont{Ciuchi}}, \bibnamefont{and}
  \bibinfo{author}{\bibfnamefont{J.~R.} \bibnamefont{Sambles}},
  \bibinfo{journal}{Phys. Rev. E} \textbf{\bibinfo{volume}{64}},
  \bibinfo{pages}{021708} (\bibinfo{year}{2001}).

\bibitem[{\citenamefont{Jewel and Sambles}(2002)}]{Sambl:jap:2002}
\bibinfo{author}{\bibfnamefont{S.~A.} \bibnamefont{Jewel}} \bibnamefont{and}
  \bibinfo{author}{\bibfnamefont{J.~R.} \bibnamefont{Sambles}},
  \bibinfo{journal}{J. Appl. Phys.} \textbf{\bibinfo{volume}{92}},
  \bibinfo{pages}{19} (\bibinfo{year}{2002}).

\end{thebibliography}

\end{document}